\documentclass{emulateapj}
\usepackage{natbib}
\usepackage{epsfig}
\usepackage{enumitem}
\usepackage{color}

\newcommand{\acounits}{\mbox{M$_\odot$ pc$^{-2}$} \mbox{(K km s$^{-1}$)$^{-1}$}}

\newcommand{\beameps}{\mbox{$\left< \epsilon \right>$}}

\shorttitle{Density Distributions and Line Ratio Patterns}
\shortauthors{Leroy et al.}

\begin{document}

\slugcomment{Accepted for publication in the Astrophysical Journal} 
\title{Millimeter-Wave Line Ratios and Sub-beam Volume Density Distributions}
\author{
Adam K. Leroy\altaffilmark{1},
Antonio Usero\altaffilmark{2},
Andreas Schruba\altaffilmark{3},
Frank Bigiel\altaffilmark{4},
J.~M.~Diederik Kruijssen\altaffilmark{5,6},
Amanda Kepley\altaffilmark{7},
Guillermo A. Blanc\altaffilmark{8,9,10},
Alberto D. Bolatto\altaffilmark{11},
Diane Cormier\altaffilmark{6},
Molly Gallagher\altaffilmark{1},
Annie Hughes\altaffilmark{12,13},
Maria J. Jim\'enez-Donaire\altaffilmark{6},
Erik Rosolowsky\altaffilmark{14}
Eva Schinnerer\altaffilmark{5}
}
\altaffiltext{1}{Department of Astronomy, The Ohio State University, 140 West 18th Avenue, Columbus, OH 43210}
\altaffiltext{2}{Observatorio Astron�mico Nacional (IGN), C/ Alfonso XII, 3, 28014 Madrid, Spain}
\altaffiltext{3}{Max-Planck-Institut f\"ur extraterrestrische Physik, Giessenbachstra{\ss}e 1, 85748 Garching, Germany}
\altaffiltext{4}{Institute f\"ur theoretische Astrophysik, Zentrum f\"ur Astronomie der Universit\"at Heidelberg, Albert-Ueberle Str. 2, 69120 Heidelberg, Germany.}
\altaffiltext{5}{Astronomisches Rechen-Institut, Zentrum f\"{u}r Astronomie der Universit\"{a}t Heidelberg, M\"{o}nchhofstra\ss e 12-14, 69120 Heidelberg, Germany}
\altaffiltext{6}{Max Planck Institute f\"ur Astronomie, K\"onigstuhl 17, 69117, Heidelberg, Germany}
\altaffiltext{7}{National Radio Astronomy Observatory, 520 Edgemont Road, Charlottesville, VA 22903, USA}
\altaffiltext{8}{Departamento de Astronom\'{\i}a, Universidad de Chile, Casilla 36-D, Santiago, Chile}
\altaffiltext{9}{Centro de Astrof\'{\i}sica y Tecnolog\'{\i}as Afines (CATA), Camino del Observatorio 1515, Las Condes, Santiago, Chile}
\altaffiltext{10}{Visiting Astronomer, Observatories of the Carnegie Institution for Science, 813 Santa Barbara St, Pasadena, CA, 91101, USA}
\altaffiltext{11}{Department of Astronomy, Laboratory for Millimeter-wave Astronomy, and Joint Space Institute, University of Maryland, College Park, Maryland 20742, USA}
\altaffiltext{12}{CNRS, IRAP, 9 av. du Colonel Roche, BP 44346, F-31028 Toulouse cedex 4, France}
\altaffiltext{13}{Universit\'{e} de Toulouse, UPS-OMP, IRAP, F-31028 Toulouse cedex 4, France}
\altaffiltext{14}{Department of Physics, University of Alberta, Edmonton, AB, Canada}

\begin{abstract}
We explore the use of mm-wave emission line ratios to trace molecular gas density when observations integrate over a wide range of volume densities within a single telescope beam. For observations targeting external galaxies, this case is unavoidable. Using a framework similar to that of Krumholz \& Thompson (2007), we model emission for a set of common extragalactic lines from lognormal and power law density distributions. We consider the median density of gas producing emission and the ability to predict density variations from observed line ratios. We emphasize line ratio variations, because these do not require knowing the absolute abundance of our tracers. Patterns of line ratio variations have the prospect to illuminate the high-end shape of the density distribution, and to capture changes in the dense gas fraction and median volume density. Our results with and without a high density power law tail differ appreciably; we highlight better knowledge of the PDF shape as an important area. We also show the implications of sub-beam density distributions for isotopologue studies targeting dense gas tracers. Differential excitation often implies a significant correction to the naive case. We provide tabulated versions of many of our results, which can be used to interpret changes in mm-wave line ratios in terms of changes in the underlying density distributions.
\end{abstract}


\section{Introduction}
\label{sec:intro}

Gas volume density, $\rho$, plays a central role in most theories explaining star formation in molecular clouds \citep[e.g., see][among many others]{GAO04,MCKEE07,LADA10,PADOAN11,FEDERRATH12,KRUMHOLZ12B}. Many models, especially those based on turbulent clouds \citep[e.g.,][]{PADOAN02,KRUMHOLZ05,HENNEBELLE11,FELDMANN11}, consider the free fall time $\tau_{\rm ff} \propto \rho^{-0.5}$ as the governing timescale for star formation, with an impact on the output cluster population \citep[e.g.,][]{KRUIJSSEN12}. Another class of observationally motivated ``threshold models'' \citep{GAO04,WU05,HEIDERMAN10,LADA10,WU10,LADA12,EVANS14}, posit that stars form only in the densest parts of clouds, with the star formation rate driven by the gas mass above a gas volume density threshold. 

This important theoretical role for gas density agrees with observations of local clouds. These show star formation confined to the highest density regions \citep{LADA03,HEIDERMAN10,LADA10,ANDRE14}, and that the amount of high column density material in a cloud correlates with its ability to form stars \citep{KAINULAINEN09}.

Testing these theories requires estimating gas density across galaxies. Such estimates also offer the prospect to understand how galaxies set the density of their gas. The nearby galaxy population offers access to a range of conditions and a clean external perspective not available in the Milky Way, and so offers many advantages as a laboratory to explore the role of gas density. The challenge to making such measurements is that the concentrations of dense gas seen in Milky Way clouds are very compact, $\sim 0.1{-}1$~pc \citep[e.g.,][]{LADA03,ANDRE14} and lie within larger, lower density superstructures. As a result, even within the $\sim 40$~pc beam of the highest resolution gas maps targeting nearby galaxies \citep[e.g.,][]{SCHINNERER13}, a vast range of gas volume densities are convolved together within an individual extragalactic beam.

Given this limitation, multi-line spectroscopy has become the standard approach to investigate the density of gas across galaxies \citep[e.g.,][]{GAO04,GARCIABURILLO12,KEPLEY14,USERO15,CHEN15,BIGIEL16}. By contrasting a line excited at high gas densities, e.g., HCN (1-0), with a line excited at nearly all gas densities, e.g., CO (1-0), one can estimate the fraction of dense gas. In a pioneering study, \citet{GAO04B} carried out a large survey of HCN in bright galaxies and used this approach to show that, to first order, the ratio of IR-to-HCN luminosity appears constant across the galaxy population, while the ratio to IR-to-CO does not. \citeauthor{GAO04} and following authors \citep[e.g.,][]{WU05,WU10,LADA12} interpreted this as evidence for a universal gas density threshold for star formation.

More recent studies of normal star-forming galaxies by \citet{GARCIABURILLO12}, \citet{USERO15}, and \citet{BIGIEL16} reveal a more complex relationship between high critical density lines, CO emission, and star formation. \citet{USERO15} and \citet{BIGIEL16} demonstrated that both HCN~(1-0)/CO~(1-0), tracing the dense gas fraction, and IR/HCN~(1-0), tracing the efficiency with which dense gas forms stars, depend on environment. These results have the same sense as those obtained contrasting local clouds with those in the Milky Way center \citep{LONGMORE13,KRUIJSSEN14B}.

So far, these extragalactic studies have mainly considered a two-phase molecular medium in which gas is either ``dense,'' and so emits HCN or HCO$^+$, or ``not dense,'' and so emits CO. This is a poor representation of the cold interstellar medium (ISM). Turbulent theories predict a wide range of densities, distributed approximately lognormally within any given cloud \citep[e.g.,][]{PADOAN02}. Milky Way observations that resolve individual clouds show a large range of column densities, implying a large range of volume densities \citep[e.g.,][]{KAINULAINEN09,ABREUVICENTE15}. The functional form of the density distribution is still debated \citep[e.g.,][]{LOMBARDI15,SCHNEIDER15}, but all observations and models agree that a wide range of densities coexist within any individual region.

Further complicating the issue, emission from high density tracers like HCN~(1-0) or HCO$^+$~(1-0) does not come exclusively from gas above some threshold collider density. The emissivity (line emission per mass) of a molecule does peak at some collider density, and that peak depends on the critical density, optical depth, and temperature of the molecule in question. But regions with lower densities can still emit; they merely do so with lower efficiency. If low density regions outnumber higher density regions, then they may contribute appreciably to, or even dominate, emission from that molecule. This ability of gas to emit below the nominal critical density (even after modification for line trapping) has been emphasized in the Galactic literature \citep[e.g.,][and references therein]{SHIRLEY15,MANGUM15} but less explored in other galaxies.

In this paper, we consider the impact of these two effects on extragalactic observations that integrate over a wide range of distributions. We model line emission from regions that contain a wide distribution of densities, and account for a realistic dependence of emissivity on collider density. To do this, we treat a cloud as an ensemble of one-zone non-equilibrium models \citep[calculated using RADEX; ][]{VANDERTAK07} that share an optical depth, and so an escape probability. This approach follows \citet{KRUMHOLZ07B}, who laid out this cloud model but focused on how density distributions influence observed star formation scaling relations \citep[see also the closely related paper by][]{NARAYANAN08}. Here, we investigate the implications of sub-beam density distributions and realistic emissivities on observed millimeter line emission.

The paper proceeds as follows. We describe our model and calculations in \S \ref{sec:method}. Then, in \S \ref{sec:onezone_results} we note several results from one-zone models that are important to interpret emission from density distributions. In \S \ref{sec:dist_results} we show results integrated over realistic density distributions. We consider the median density producing emission, the pattern of line ratio changes induced by changing sub-grid volume density distributions, and the ability of line ratio variations to gauge changes in the dense gas mass fraction and median density. We also consider the implications of sub-beam density distributions for isotopologue ratios and dense gas conversion factors. We summarize our conclusions and discuss implications and future directions in \S \ref{sec:discuss}.

\subsection{Definitions of Densities}
\label{sec:defs}

We discuss a number of densities throughout the paper. For clarity, we summarize these here. All densities refer to molecular hydrogen, H$_2$. We neither account for helium nor consider collisions with electrons or atomic hydrogen. Given that we focus on relative statements and cold, dark gas, these approximations should have minimal impact.

\begin{itemize}
\item {\em Collider density ($n_{\rm H2}$)}: the volume density of H$_2$ molecules. This is an input to our one zone models. When we refer to density distributions, we mean collider density distributions.

\item {\em Critical density:} For a given transition and kinetic temperature, the collider density at which collisional de-excitations balance radiative de-excitations, taking no account of line trapping. See \citet{SHIRLEY15} for a recent review.

\item {\em Effective critical density}: For a given transition and kinetic temperature, the collider density at which collisional de-excitations balance radiative de-excitations, taking into account radiative line trapping. See \citet{SHIRLEY15} for a recent review.

\item {\em Most effective density for emission ($n_{\rm eff}$)}: For a given transition, kinetic temperature, and optical depth, the minimum density at which the emissivity of the line (defined below) reaches 95\% of its peak value. This is a new quantity defined in this paper. It is similar, but not identical, to the effective critical density. Often, effective critical density is used as a short-hand for the density at which gas is best at emitting. This quantity measures this directly. We present tabulated results $n_{\rm eff}$ below.

\item {\em Median density for emission ($n_{\rm med}^{\rm emis}$)}: For a given transition, kinetic temperature, optical depth, and density distribution, the collider density below which half of the line emission is generated. A closely related quantity is the main focus of \citet{KRUMHOLZ07B}.

\item {\em Median density by mass ($n_{\rm med}^{\rm mass}$)}: For a given density distribution, the collider density below which half the mass lies.
\end{itemize}

We also refer to the dense gas fraction, $f_{\rm dense}$. For a given density distribution, this is the fraction of mass in the distribution that lies above a density threshold, $n_{\rm thresh}$. We adopt $n_{\rm thresh} = 10^{4.5}$~cm$^{-3}$ by default, but also consider a range of possible threshold values.

\section{Method}
\label{sec:method}

\begin{deluxetable}{lcccc}
\tabletypesize{\scriptsize}
\tablecaption{Molecular Transitions Considered \label{tab:lines}}
\tablewidth{0pt}
\tablehead{
\colhead{Transition} &
\colhead{Frequency} & 
\colhead{$\tau$\tablenotemark{a}} &
\colhead{$n_{\rm H2} (\epsilon_{\rm max})$\tablenotemark{b}} &
\colhead{$X_{\rm mol}$\tablenotemark{c}}
\\
\colhead{} &
\colhead{(GHz)} & 
\colhead{} &
\colhead{(cm$^{-3}$)} & 
\colhead{}
}
\startdata
HCN\tablenotemark{d} $J=1\rightarrow0$ & $88.630$ & 1 & $2 \times 10^5$ & $10^{-8}$\\
HCO$^+$ $J=1\rightarrow0$ & $89.189$ & 1 & $4 \times 10^4$ & $10^{-8}$\\
HNC $J=1\rightarrow0$ & $90.663$ & 1 & $1 \times 10^5$ & $10^{-8}$ \\
CS $J=2\rightarrow1$ & $97.980$ & 1 & $7 \times 10^4$ & $10^{-8}$ \\
$^{13}$CO\tablenotemark{e} $J=1\rightarrow0$ & $110.201$ & 0.1 & $8 \times 10^2$ & $2\times 10^{-6}$ \\
$^{12}$CO $J=1\rightarrow0$ & $115.271$ & 10 & $1 \times 10^2$ & $10^{-4}$ \\
$^{12}$CO $J=2\rightarrow1$ & $230.538$ & *\tablenotemark{f} & $1 \times 10^3$ & $10^{-4}$ \\
$^{12}$CO $J=3\rightarrow2$ & $345.796$ & *\tablenotemark{f} & $9 \times 10^3$ & $10^{-4}$ 
\enddata
\tablenotetext{a}{Representative optical depth assumed for this line. When not otherwise noted, this $\tau$ is used throughout the paper.}
\tablenotetext{b}{Minimum collider density, $n_{\rm H2}$, at which the emissivity reaches 95\% of its peak value for $T=25$~K and the representative $\tau$.}
\tablenotetext{c}{Fiducial abundance of the molecule adopted in this paper. This divides out of many aspects of the analysis.}
\tablenotetext{d}{We ignore the splitting of the HCN 1-0 line because the component spacing is small compared to typical extragalactic line widths.}
\tablenotetext{e}{Results for C$^{18}$O are essentially equivalent to those for $^{13}$CO if both lines are taken to be optically thin. We plot only $^{13}$CO for clarity.}
\tablenotetext{f}{For internal consistency, we calculate emission from CO~(2-1) and CO~(3-2) using the $N / \Delta v$ calculated for CO~(1-0). Thus these lines are pinned to an assumed $\tau$ for CO~(1-0).}
\tablecomments{Data for models for these species taken from Leiden Atomic Molecular Database \citep[LAMBDA; ][]{SCHOIER05}.}
\end{deluxetable}

We consider a suite of commonly observed rotational transitions of molecules in the $80{-}115$~GHz range. These lines, listed in Table \ref{tab:lines}, are bright enough that they can be surveyed in other galaxies. They are excited at a range of densities, so that ratios among them offer the prospect to constrain changing density distributions. Focusing on the $80$--$115$~GHz range, we aim to de-emphasize excitation concerns due to temperature, highlighting the effect of changing density. We do include the $J=2\rightarrow1$ and $J=3\rightarrow2$ transitions of CO, because these are observed across many nearby galaxies and often serve as bulk gas tracers for practical reasons \citep[e.g.,][]{LEROY09,WILSON12}. This suite of lines closely resembles that mapped by the IRAM Large Program EMPIRE \citep{BIGIEL16}, the survey of \citet{USERO15}, and ALMA mapping by Gallagher et al. (in prep.). A goal of this paper is to help inform interpretation of those and similar observations.

For each molecule, we create a grid of one zone models (\S \ref{sec:onezone}), each describing a single set of physical conditions. From each, we calculate the emissivity, $\epsilon$, which will be our figure of merit for much of the paper (\S \ref{sec:emisdef}). We combine these one zone models to simulate emission from a realistic distribution of densities. To do this, we follow \citet{KRUMHOLZ07B} and make the simplifying assumption that a single escape probability, and so a single optical depth, describes each transition throughout the beam (\S \ref{sec:model}). We combine this assumption with an assumed density distribution (\S \ref{sec:distdef}) to calculate the beam-averaged emissivity, $\beameps$. Ratios of \beameps\ for different lines but otherwise matched conditions can be observed as line ratios. Thus, this approach allows us to explore the impact of changing density distributions on observed line ratios. While useful and a large improvement over a one zone treatment, this model has several shortcomings. We describe these and note directions for future improvement in \S \ref{sec:limits}.

\subsection{Grid of One Zone Models}
\label{sec:onezone}

For each molecule, we use the RADEX code \citep{VANDERTAK07} to predict the emission from a series of one zone models. Each model assumes a single collider (H$_2$) volume density, $n_\mathrm{H2}$, kinetic temperature, $T_{\rm kin}$, and column-per-line width, $N_{\rm mol} / \Delta v$, of the molecule\footnote{We treat the column density (of the observed molecule) per unit line width, $N_{\rm mol} / \Delta v$, rather than $N_{\rm mol}$ and $\Delta v$ separately. In the escape probability treatment, it is this quantity that maps to optical depth rather than the two quantities separately.}. Given these inputs, RADEX solves for the level populations and predicted emission without assuming local thermodynamic equilibrium (LTE).

Our model grid covers a range of kinetic temperatures, $T_{\rm kin} = 10{-}300$~K, and a range of H$_2$ collider densities, $n_{\rm H2} = 10^1{-}10^8$~cm$^{-3}$. This range of collider (H$_2$) densities spans from well below to well above the critical density of each transition of interest. The range of column-per-line width, $N / \Delta v$, is $10^{13}{-}10^{19}$~cm$^{-2}$~(km~s$^{-1}$)$^{-1}$ for the CO molecules and $10^{10}{-}10^{16}$~cm$^{-2}$~(km~s$^{-1}$)$^{-1}$ for the other molecules. We chose these ranges to span from optically thin ($\tau \lesssim 0.1$) to optically thick ($\tau \gtrsim 10$) for each molecule. 

We space grid points logarithmically along each axis of the grid, sampling 60 points along the $T_{\rm kin}$ and $N_{\rm mol} / \Delta v$ axes and 70 points along the $n_{\rm H2}$ axis. At each $\left( n_{\rm H2}, T_{\rm kin}, N_{\rm mol}/\Delta v \right)$ point, we record the intensity, $I$, of each low $J$ line in K~km~s$^{-1}$, and its optical depth, $\tau$. The result is a grid of $\tau (n_{\rm H2}, T_{\rm kin}, N_{\rm mol}/\Delta v)$ and $I (n_{\rm H2}, T_{\rm kin}, N_{\rm mol}/\Delta v)$ that serves as the backbone of our calculations.

All models assume the expanding sphere (classic large velocity gradient, LVG) geometry \citep{GOLDREICH74,SOBOLEV60}. We use the version of RADEX released on November 30, 2011 and adopt molecular data from the Leiden Atomic Molecular Database \citep[LAMBDA; ][]{SCHOIER05} for HCN, HNC, HCO$^+$, CS, $^{13}$CO, $^{12}$CO. We also ran model grids for C$^{18}$O. The results are very similar to those for $^{13}$CO and we only present the $^{13}$CO results here. To use our model for C$^{18}$O, we suggest to divide the emissivity relative to H$_2$ by the $^{13}$CO/C$^{18}$O abundance ratio.

\subsection{Emissivity}
\label{sec:emisdef}

For each one zone model, we calculate the emissivity, $\epsilon$, of each transition. We define emissivity as the intensity, $I$, divided by the column density, $N$,

\begin{equation}
\epsilon = \frac{I}{N}~.
\end{equation}

\noindent Thus, the emissivity, $\epsilon$, measures how effectively gas emits in the specified transition for the conditions in the model. It has units of K~km~s$^{-1}$~(cm$^{-2}$)$^{-1}$. Along with a mean molecular mass, the emissivity can be recast to have units of K~km~s$^{-1}$~(M$_\odot$~pc$^{-2}$)$^{-1}$. Therefore $\epsilon$ is the mass-to-light ratio of gas with the given $(n_{\rm H2}, T_{\rm kin}, N_{\rm mol}/\Delta v)$.

In practice, we are interested in the emitted intensity per unit total gas (H$_2$) rather than emitted intensity per unit mass of HCN or CO. Therefore, our calculations use the column of the emitting molecule, $N_{\rm mol}$.  Thus, when we write $\epsilon$ below we mean

\begin{equation}
\label{eq:emis}
\epsilon = \frac{I}{N_{\rm H2}} = \frac{I}{N_{\rm mol}~X ({\rm mol})^{-1}}~.
\end{equation}

\noindent Here $X ({\rm mol})$ refers to the abundance of the molecule, with $X ({\rm mol}) \equiv N_{\rm mol} / N_{\rm H2}$. Our models yield $I / N_{\rm mol}$. We note our fiducial values of $X ({\rm mol})$ for each molecule in Table \ref{tab:lines}. These are $10^{-8}$ for CS, HCN, HCO$^+$, HNC, $10^{-4}$ for CO, and $2 \times 10^{-6}$ for $^{13}$CO (i.e., 50 times lower than CO). These are motivated by existing abundance estimates \cite[e.g.,][]{MARTIN06}, but for many of the results in this paper, the exact values of $X ({\rm mol})$ are not important. Instead, we will be interested in how the ability of molecular gas to emit in these lines varies as a function of changing density distributions, optical depth, and so on. As discussed in the next section, our model treatment implies modest changes in at least some values of $X ({\rm mol})$ within the model framework.

\subsection{Distributions of Densities With Shared $\tau$ and $T_{\rm kin}$} 
\label{sec:model}

Observations of a whole cloud or a large part of a galaxy will blend emission from gas at many densities. To simulate this, we combine one zone models, which each have a single associated $n_{\rm H2}$, to model a distribution of densities. Our key simplifying assumption is that the emission for each line comes from gas that shares a single $T_{\rm kin}$ and $\tau$. That is, we consider a distribution of densities, but take the gas to be isothermal and adopt a fixed optical depth within each beam.

The condition of fixed $\tau$, and so fixed escape probability, follows \citet{KRUMHOLZ07B}, though our implementation differs in detail. For a given transition, $T_{\rm kin}$, and $n_{\rm H2}$, $\tau$ depends on $N_{\rm mol} / \Delta v$. While column densities at the scales relevant to radiative transfer can be difficult to gauge at extragalactic distances, $\tau$ can be accessed by observations, including via comparison of rarer isotopologues to the main gas tracers (e.g., Jiminez-Donaire et al.\ MNRAS submitted).

For the rest of the paper, we will not focus on the actual values of the column density, $N_{\rm mol}$, or column-per-line width, $N_{\rm mol} / \Delta v$. Instead, for each transition we will consider $\epsilon (n_{\rm H2}, T_{\rm kin}, \tau)$, where $\tau$ is the optical depth of that transition. We derive this quantity from our grid of one zone models by selecting the appropriate $n_{\rm H2}$ and $T_{\rm kin}$ and then interpolating along the curve of $N_{\rm mol} / \Delta v$ vs. $\tau$ for our model grid to select $N_{\rm mol} / \Delta v$ that gives the desired value of $\tau$. Then we interpolate within our grid to calculate $\epsilon (n_{\rm H2}, T_{\rm kin}, \tau)$. Because of the fine spacing of the model grid, the interpolation has only a minor influence on the results.

Fixing $T_{\rm kin}$ and $\tau$, we consider a distribution of volume densities, $P(n_{\rm H2})$. We sum over all densities to derive the beam averaged emissivity:

\begin{equation}
\label{eq:beamemis}
\left< \epsilon \right> = \frac{\int n_{\rm H2}~P(n_{\rm H2})~\epsilon (n_{\rm H2}, T, \tau) dn_{\rm H2}}{\int n_{\rm H2}~P(n_{\rm H2})~dn_{\rm H2}}~.
\end{equation}

\noindent The beam-averaged emissivity, $\beameps$, resembles the one zone emissivity (Equation \ref{eq:emis}) in that it describes the intensity of the line per column of H$_2$. However, \beameps\ now measures the effective emissivity of a whole distribution of densities convolved together.

We weight by $n_{\rm H2}$ in Equation \ref{eq:beamemis} because $P(n_{\rm H2})$ represents the probability of finding a gas volume density $n_{\rm H2}$ in a given volumetric cell within the cloud. The amount of emission from gas at $n_{\rm H2}$ will trace the total amount of mass at that density, rather than the amount of volume. The amount of mass at $n_{\rm H2}$ is given by $n_{\rm H2} P(n_{\rm H2})$.

{\em Multiple Transitions of the Same Molecule:} Picking $\tau$ for one transition while fixing $n_{\rm H2}$ and $T_{\rm kin}$ implies $N_{\rm mol} / \Delta v$. With $n_{\rm H2}$,  $T_{\rm kin}$, and $N_{\rm mol} / \Delta v$ specified, the emission from other transitions of the same molecule is also determined. Thus, once we specify $\tau$ for CO~(1-0), the optical depth of CO~(2-1) and CO~(3-2) are no longer free parameters. In this paper, this issue arises mainly for CO. In general, our approach is to specify an observationally motivated $\tau$ for a single transition for each molecule. We then calculate emission from the other transitions self-consistently \citep[see][for a similar approach treating Galactic observations of ammonia]{ROSOLOWSKY08C}. This means that the optical depth of these other transitions may not be fixed across the model distribution. In exchange, the calculated line ratios among multiple transitions of a single molecule will make physical sense.

{\em Implicit Changes in Relative Abundances:} A corollary of this point is that our model implies modest changes in the relative abundance of the molecules that we treat. For each molecule at each density, we pick $N_{\rm mol}/\Delta v$ to yield our fiducial $\tau$ for the specified transition. Because we fix the distribution of $n_{\rm H2}$, the ratio $N_{\rm mol}/n_{\rm H2}$ will change from density to density to keep $\tau$ fixed. As a result, the relative values of, e.g., $N_{\rm HCN}$ and $N_{\rm CO}$ will vary within a model distribution. This does not represent an inherent inconsistency, but does assume that mild abundance variations will conspire to yield fixed $\tau$. We discuss alternative formulations below (\S \ref{sec:limits}). The \citet{KRUMHOLZ07B} model suffers from the same internal inconsistency. They formulate this as a cloud radius that varies from transition to transition, so that in their model $N_{\rm mol} / \Delta v$ can vary because $N_{\rm mol} \approx n_{\rm H2}~X ({\rm mol}) / R_{\rm mol}$ varies with $R_{\rm mol}$, where $R_{\rm mol}$ is the radius of the emitting region. Because radius of the HCN-emitting region, $R_{\rm HCN}$, and the CO-emitting region, $R_{\rm CO}$, can differ, the total abundance will also vary in this formulation.

\subsection{Realistic Density Distributions} 
\label{sec:distdef}

\begin{figure}[t]
\plotone{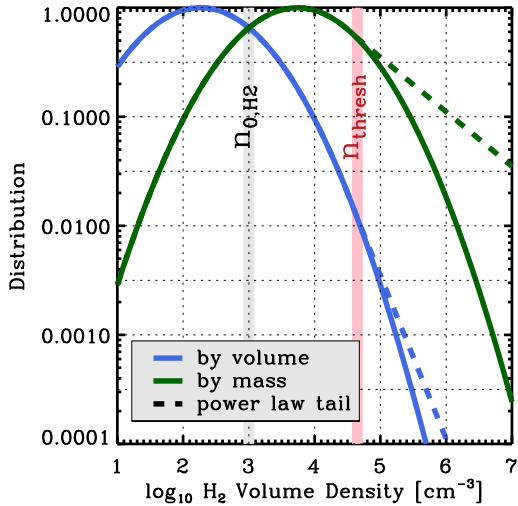}
\caption{Illustration of model density distributions by volume (blue) and mass (green), with (dashed) and without (solid) power law tails at high densities. Vertical lines indicate the mean density (gray) and the adopted threshold for the onset of a power law tail (pink), when one is present. Although a large part of the volume holds low density gas, most of the mass resides in a small part of the volume with high densities \citep[see][]{PADOAN02}. Similarly, although the power law tail adds only a small amount of additional volume at high densities, it contains ${\sim} 20\%$ of the mass for this model.}
\label{fig:pdfs}
\end{figure}

We explore emission from lognormal density distributions, sometimes modified to include a power law tail at high density. Lognormal distributions are expected based on simulations \citep[e.g.,][]{VAZQUEZSEMADENI94,MACLOW04,ELMEGREEN04}, and we follow the description outlined for turbulent gas in \citet{PADOAN02}. Observations do support the existence of lognormal column density distributions within a cloud \citep[e.g.,][]{KAINULAINEN09,RATHBORNE14,ABREUVICENTE15}. However, power law descriptions of the column density distribution may be equally valid given observational uncertainties \citep[e.g.,][]{LOMBARDI15}. Even when a lognormal describes the bulk of the gas in a cloud, self-gravity may lead the densest gas to exhibit power-law density distributions \citep[e.g.,][]{KAINULAINEN09,KAINULAINEN13,STUTZ15}. 

Note that the appropriate form of the density distribution on scales larger than a cloud, including the $\gtrsim$kpc regions observed by many mm-wave spectral studies, remains a subject of investigation. On larger (60~pc) scales, molecular gas surface densities do appear to exhibit an approximately lognormal distribution \citep{LEROY16}, though processes other than turbulence may create the lognormal surface density distributions observed at these large scales \citep{BERKHUIJSEN15}. Simulations show density distributions that can depend strongly on environment and the nature and strength of stellar feedback \citep[e.g.][]{HOPKINS13B}. At the least, one might expect the superposition of a diverse population of lognormal and power law distributions. Here we begin the the simplest assumption and note further explanation as an area for future work.

{\em Lognormal Distributions:} In the \citet{PADOAN02} formulation, the volume elements in a cloud show a distribution of gas volume densities described by:

\begin{equation}
\label{eq:lognormal}
d P (\ln n^\prime)  \propto \exp \left( - \frac{(\ln n^\prime - \overline{\ln n^\prime})^2}{2 \sigma^2}  \right) d \ln n^\prime~,
\end{equation}

\noindent where $dP$ is the fraction of cells with volume densities in a logarithmic step $d \ln n^\prime$ centered on $n^\prime$; $n^\prime = n_{\rm H2} / n_0$ is the volume density normalized by the mean volume density, $n_0$; and $\sigma$ is the width of the distribution. The distribution does not peak at the mean volume density, $\overline{n^\prime} \equiv 1$, but at the mean of the logarithm, $\overline{\ln n^\prime}$. The two are related via $\overline{\ln n^\prime} = -0.5 \sigma^2$ \citep[see][for details]{PADOAN02}. In practice, we work in $\log_{10}$ rather than $\ln$, so the $\sigma$ that we quote are in dex and differ from those in the \citet{PADOAN02} formalism by a factor of $\ln 10 \approx 2.3$.

Equation \ref{eq:lognormal} describes the distribution of volume densities in the cloud. This differs from the distribution of mass because higher density cells hold more mass. The fraction of mass in a given logarithmic density step is $dm \propto n dP$. Thus, as discussed by \citet{PADOAN02}, most mass resides in dense substructures that occupy a relatively small fraction of the volume.

In practice, we implement Equation \ref{eq:lognormal} numerically, also creating a mass distribution from $dm = n dP$. Each lognormal distribution is specified by two numbers: the mean density, $n_0$, and the rms logarithmic width, bookkept in dex (i.e., $\log_{10}$), $\sigma$. Following \citet{PADOAN02}, for a distribution resulting from isothermal supersonic turbulence, $\sigma$ is closely related to the three dimensional turbulent Mach number, $\mathcal{M}$. In that case

\begin{equation}
\sigma \approx 0.43~\sqrt{\ln (1+0.25 \mathcal{M}^2)}~{\rm dex}~.
\end{equation}

\noindent Typical Mach numbers in spiral and starburst galaxies span the range $\mathcal{M} \approx 5{-}100$, so that we will mainly be concerned with $\sigma \approx 0.6{-}1.2$~dex \citep[e.g., see][]{LEROY16}. Unless otherwise noted, we adopt a fiducial $\sigma = 0.8$~dex, corresponding to $\mathcal{M} \sim 10$.

{\em Power Law Tails:} Observations and simulations suggest that self-gravity in the densest parts of a cloud leads to the formation of a power law tail in $P(n)$ at high $n_{\rm H2}$. This effect is observed to vary from cloud-to-cloud, with the amount of mass in the tail correlating with the star-forming activity \citep{KAINULAINEN09,ABREUVICENTE15}.

We explore this effect by incorporating a power law tail into some of our density distributions. In these cases, we follow \citet{FEDERRATH13} and take $P(n_{\rm H2})$ to be a power law of form

\begin{equation}
\label{eq:powerlaw}
d P (\ln n^\prime)  \propto \exp \left( \alpha \ln n^\prime \right)~{\rm where}~n^\prime > n^\prime_{\rm thresh}~.
\end{equation}

\noindent Here $n^\prime_{\rm thresh}$ is the threshold for the onset of the power law tail in units of the density normalized by the mean, $n_0$. $\alpha$ is the slope of the power law in logarithmic units. Note that this is offset by one power from the slope in non-logarithmic units, so that $dP(n) \propto n^{\alpha - 1}$ \citep[see][]{FEDERRATH13}. As a result, for any $\alpha < -1$ the mass distribution ($\propto \int n \times n^{\alpha-1} dn$) will converge. In these models, a lognormal distribution still describes gas at $n^\prime < n^\prime_{\rm thresh}$.

Following \citet{FEDERRATH13} and \citet{VALLINI16}, we take $\ln n^\prime_{\rm thresh} \approx 3.8$. That is, we set the power law to begin at $\sim 45$ times the mean density of the cloud. We take $\alpha = -1.5$, intermediate in the range of $-1$ to $-2.5$ found by \citet{FEDERRATH13}, though towards the steep end of the values observed for column density distribution in Orion \citep{STUTZ15}. In principle, the appropriate threshold for the onset of self-gravity depends on the Mach Number and other quantities \citep[see][]{FEDERRATH13}, with our adopted value appropriate for $\mathcal{M} \approx 7$ \citep{KRUMHOLZ05,PADOAN11}. Parameter studies of $n_{\rm thresh}^\prime$ and $\alpha$ will be useful but lie beyond the scope of this paper.

Figure \ref{fig:pdfs} illustrates our model distributions of volume densities for an example case with mean density $n_{\rm 0, H2} = 10^3$~cm$^{-3}$ and width $\sigma = 0.8$~dex. We show both the amount of volume and the amount of mass at each density, and illustrate the distribution with and without a power law tail. In this case, the power law tail integrated from $n^\prime_{\rm thresh}$ through $n_{\rm H2} = 10^7$~cm$^{-3}$ holds ${\sim} 20\%$ of the total mass. The same density range in the pure lognormal case holds only ${\sim} 7\%$ of the mass. Thus, despite the modest fraction of volume in the power law tail, it has an appreciable impact on the mass distribution.

\subsection{Emission From Distributions of Densities}

We create density distributions for a range of $\sigma$ and $n_0$ appropriate to normal spiral galaxies and modest starbursts. In each case, we make two versions of the distribution: one with a power law tail and one without a power law tail. From these density distributions, we calculate the corresponding mass distributions. We note the volume density below which 50\% of the mass lies and the fraction of the mass in the power law tail, if one is implemented. Finally, we calculate the fraction of the mass that lies at densities above $n_{\rm H2} = 10^{4.5}$~cm$^{-3}$, a value commonly conflated with ``dense'' gas.

We then combine our mass distributions with our grid of one zone models, assuming a fixed $T_{\rm kin}$ and $\tau$. For each model and each line in Table \ref{tab:lines}, we calculate \beameps, as defined in Equation \ref{eq:beamemis}. This is the emission per unit mass of the whole observed region. We also record the volume density below which 50\% of the light is emitted.

The tables present results from these calculations, which serve as the basis for the plots and discussion in the rest of this paper.

\subsection{Limitations of the Model and Directions for Improvement}
\label{sec:limits}

Our major simplifying assumption is that all zones within a beam share the same $\tau$. That is, our model has a vastly simplified geometry. As mentioned above, fixing $\tau$ imposes modest variations in the relative abundances of different species. These also exist in the \citet{KRUMHOLZ07B} model, manifesting as different cloud radii for each molecule. 

An alternative approach would be to specify $\tau$ for a single transition and a single species (e.g., CO) and then adopt a fixed abundance pattern to derive $N_{\rm mol} / \Delta v$ for all other species, allowing their optical depth to vary. We consider assuming fixed $\tau$ to impose moderate optical depth on the key transitions \citep[something apparently required by observations, e.g.,][Jimenez-Donaire et al. submitted]{MEIER15} more reliable than adopting an uncertain abundance pattern.

One can imagine alternative configurations, with coupling between adjacent zones or built-in correlations, e.g., between $\tau$ and $n_{\rm H2}$. A natural alternative to fixed $\tau$ would be to impose some dynamical state on the individual zones and so link $n_{\rm H2}$ to $N/\Delta v$. Such models have been frequently invoked in the context of the CO-to-H$_2$ conversion factor \citep[e.g.,][]{MALONEY88,BOLATTO13A} and used to treat ensembles of clouds \citep[e.g.,][]{AALTO15}. The validity of imposing such a constraint on each small range of gas densities within a cloud is unclear, but this represents a direction for exploration. In particular, assuming a dynamical state for the power law tail, when present, represents a natural addition to the model.

We assume that a single, constant temperature, $T_{\rm kin}$, describes our clouds, and consider a restricted range, $T_{\rm kin} \approx 15{-}35$~K for most of the paper, taking $T_{\rm kin}=25$~K by default. The model grid includes a larger range of temperatures. Although these are not the focus of the paper, accounting for the effects of $T_{\rm kin}$ variations will be key to explore with the contrast between central molecular zones and galaxy disks. This contrast between these two common environments is dramatic and readily observed in both the Milky Way \citep[e.g.,][]{LONGMORE13,KRUIJSSEN14B} and other galaxies \citep[e.g.,][]{USERO15}. Density is thought to play a key role in this contrast \citep[e.g.,][]{KRUMHOLZ15,KRUMHOLZ16}, but temperature variations will also need to be accounted for. More, resolved Galactic clouds do show significant temperature substructure, and a density-temperature or optical depth-temperature correlation could be added to the model in place of fixed $T_{\rm kin}$. However, at present we have no general prescription to implement.

Finally, beyond the variations needed to fix $\tau$, we assume fixed abundances, $X ({\rm mol})$, throughout our models. One could modify Equations \ref{eq:emis} and \ref{eq:beamemis} to account for varying abundances. These could change within the distribution as a function of density or appear as overall changes in the normalization of \beameps\ between measurements. Implementing such chemical variations is beyond the scope of this paper, but even simple PDR models show significant chemical substructure within clouds, so this is a productive direction for future investigation. Here, we begin using the simplest approach, and trust that the strongest abundance variations occur at the lowest and highest densities, while most of the mass, and the focus of this study, rests at intermediate scales.

To address these concerns, comparison to PDR models \citep[e.g.,][]{LEBOURLOT12}, numerical simulations that can capture complex cloud geometry \citep[e.g.,][]{GLOVER12}, and detailed observations of Galactic clouds offer a clear way forward. The DESPOTIC code by \citet{KRUMHOLZ14} offers the prospect to make several next steps; it allows multi-zone clouds and implements some basic PDR structure.

For the remainder of the paper, we emphasize that insight can be gained from our simple approach, which already offers a large improvement over one zone models.

\section{Results}

\begin{figure*}
\plotone{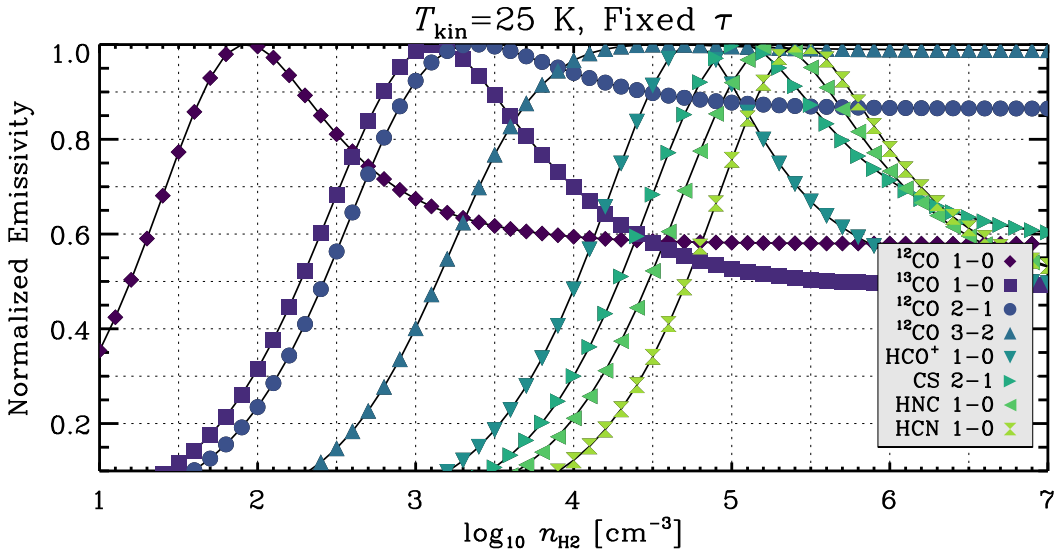}
\caption{Visualization of our one zone model grid. Normalized emissivity as a function of collider volume density, $n_{\rm H2}$, for our considered transitions at their nominal optical depths (Table \ref{tab:lines}; the curves in the figure match the order of the legend from left to right). Here, normalized emissivity is defined as emission per unit column density divided by the maximum for that line at any density. The suite of lines that we consider is excited at a range of volume densities, so that observing all of these lines in a beam gives sensitivity to the distribution of volume densities within the beam. The dependence of emissivity on volume density is not a perfect step function. Instead material at densities below the nominal effective density still emits, only somewhat less efficiently.}
\label{fig:emis_vs_dens}
\end{figure*}

Our models predict how mm-wave line emission depends on changes in the sub-beam density distribution, and so give a framework to explore density variations across galaxies. Before discussing these results, we show the dependence of emissivity, $\epsilon$, on $n_{\rm H2}$ and $\tau$ in our grid of one zone models (\S \ref{sec:onezone_results}). We then consider emission from distributions of densities (\S \ref{sec:dist_results}). We examine the median density from which emission emerges (\S \ref{sec:med_dens}), the emergent line ratio pattern (\S \ref{sec:ratios}), and the ability to estimate changes in the density distribution based on line ratio variations (\S \ref{sec:deltas}). We also use our models to explore the generality of dense gas conversion factors (\S \ref{sec:alpha}) and complications in the use of optically thin isotopologues to gauge optical depths (\S \ref{sec:isotopes}). 

\subsection{Emissivity, Density, and Optical Depth in One Zone Models}
\label{sec:onezone_results}

\begin{figure*}
\plottwo{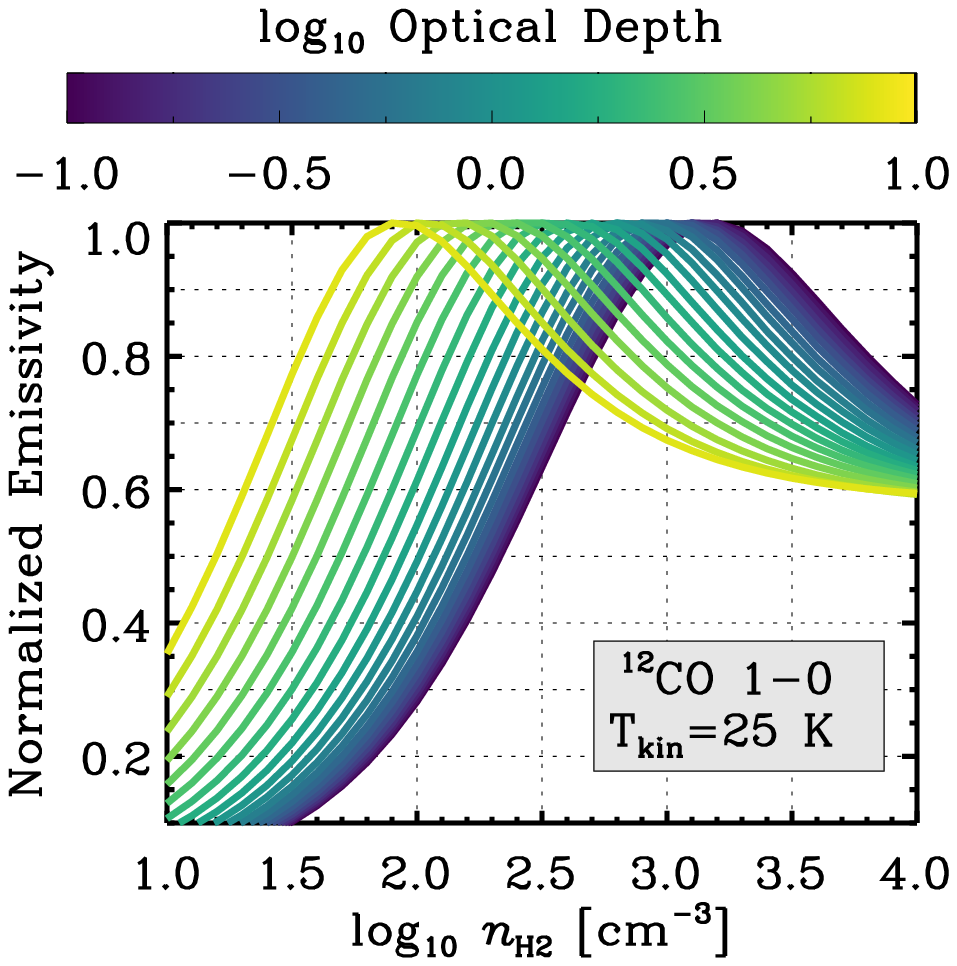}{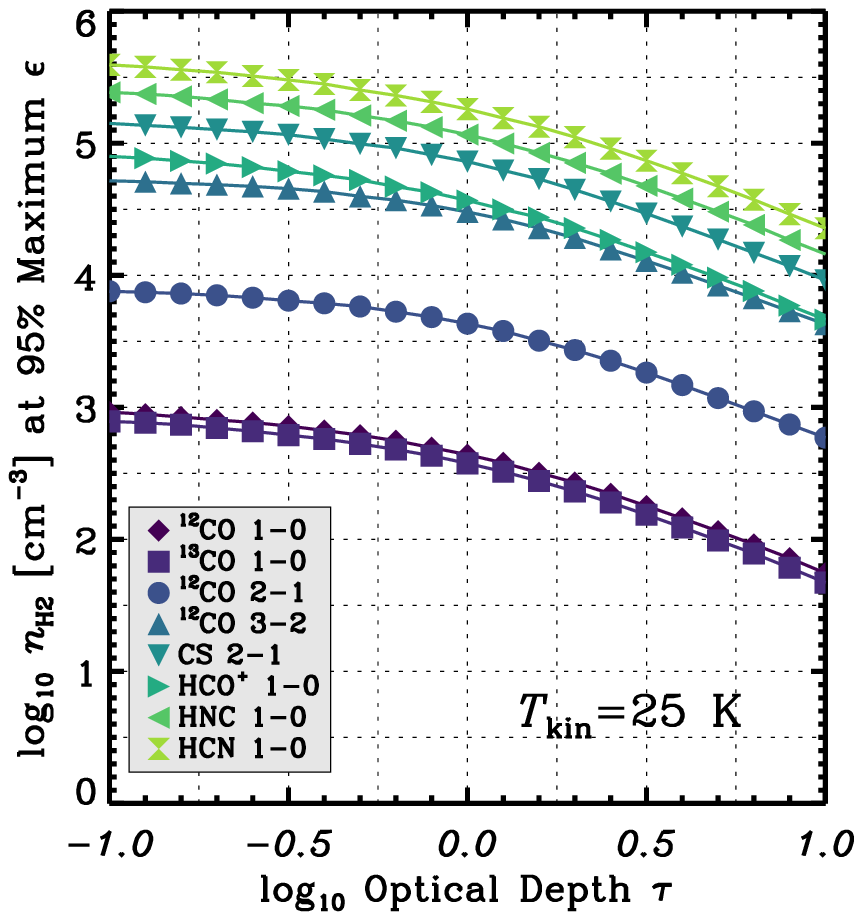}
\caption{Visualization of our one zone model grid. ({\em left}) Normalized emissivity ($y$-axis) as a function of collider volume density ($x$-axis) for $^{12}$CO~(1-0) considering a range of optical depths, $\tau$ (color). Lines with lower optical depths (blue) require higher collider densities to reach their maximum emissivity. The result is that thin CO transitions, like $^{13}$CO (1-0) or C$^{18}$O (1-0), emit maximally well at higher densities than thick transitions. This is the well-known effect of ``line trapping'' lowering the effective critical density of a transition \citep[e.g., see][]{GOLDREICH74,SHIRLEY15}. ({\em right}) Most effective density for emission from one zone non-LTE RADEX models as a function of assumed optical depth for our considered transitions over the range $\tau=0.1$ to $\tau=1$ (the lines proceed in order of the legend from bottom to top of the plot). At the low end, this approaches the classical critical density. At the high end, the standard factor of $\tau^{-1}$ to account for line trapping applies.}
\label{fig:neff_vs_tau}
\end{figure*}

\begin{figure*}
\plottwo{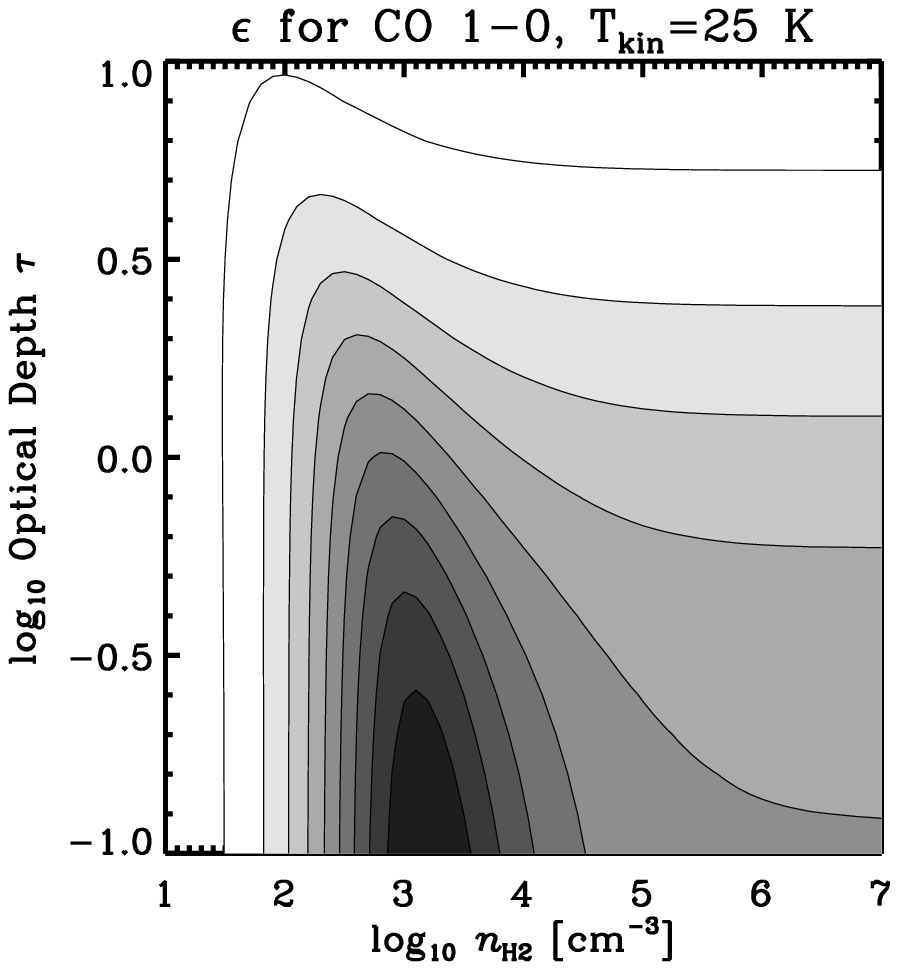}{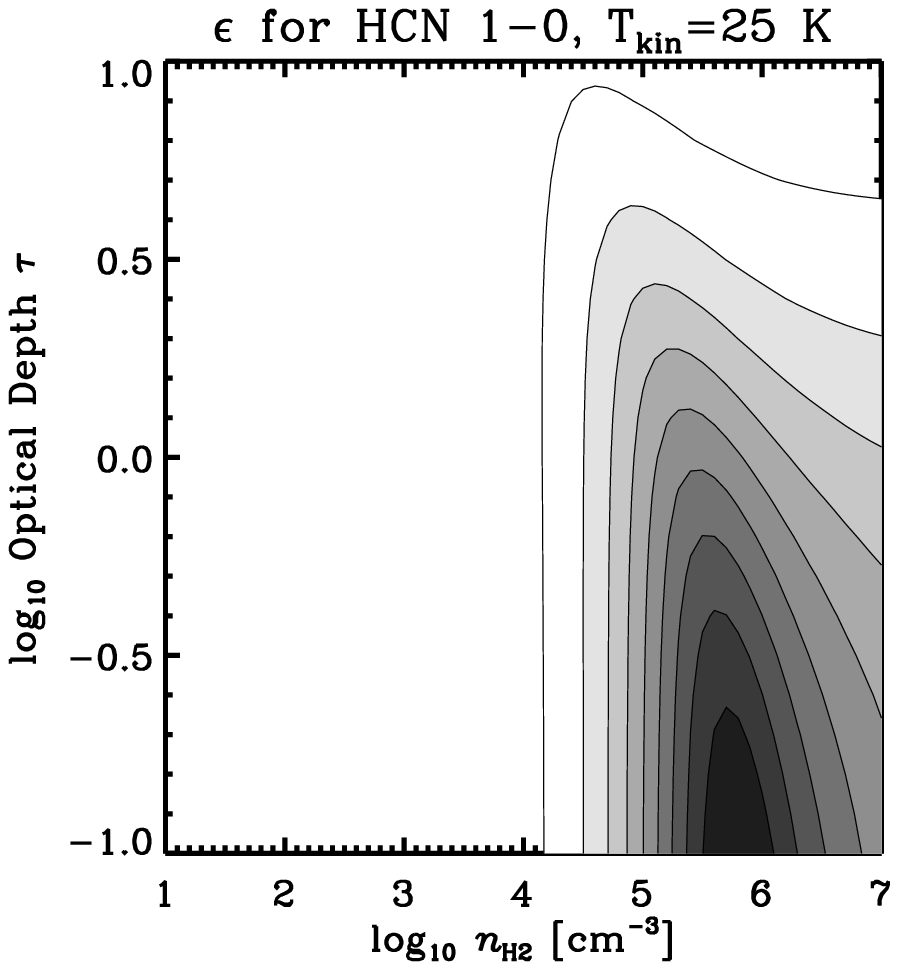}
\caption{Visualization of our one zone model grid. Emissivity, $\epsilon$, of CO ({\em left}) and HCN ({\em right}) at fixed $T_{\rm kin}$ as a function of optical depth, $\tau$, and collider density, $n_{\rm H2}$. Contours show 10, 20, 30, $\ldots$, 90\% of the peak $\epsilon$ over the whole plane. $T_{\rm kin} = 25$~K for all models. The plots show decrease of $n_{\rm H2}$ for maximum $\epsilon$ with increasing $\tau$ and the decrease in $\epsilon$ with increasing $\tau$ at fixed $n_{\rm H2}$.}
\label{fig:emis_twod}
\end{figure*}

The emissivity, $\epsilon$, varies across our grid of one zone models as a function of $n_{\rm H2}$, $T_{\rm kin}$, and $\tau$. The variation of $\epsilon$ with $n_{\rm H2}$ will be key to understand our results for density distributions. Figure \ref{fig:emis_vs_dens} shows $\epsilon (n_{\rm H2})$ for fixed $T_{\rm kin} = 25$~K and the $\tau$ in Table \ref{tab:lines}. The absolute emissivity of the molecules varies from line to line. To highlight the dependence on $n_{\rm H2}$, we normalize results for each line by the maximum $\epsilon$ for any density in the calculation. 

Figure \ref{fig:emis_vs_dens} illustrates the sensitivity of our line suite to a wide range of collider densities. It also shows that while $\epsilon$ does peak at some $n_{\rm H2}$ value, it can be substantial even at much lower densities. In this calculation, $\epsilon$ remains $\gtrsim 0.1$ times its peak value even at $n_{\rm H2}$ ten times lower than the density for peak emissivity. That is, molecules still emit effectively well below the density that maximizes their emissivity. This point has been recognized and widely emphasized in the Galactic literature \citep[see][]{EVANS99,SHIRLEY15}. In extragalactic observations, where observations necessarily average over a wide range of densities in a single beam, it is potentially even more important than in the Milky Way.

For the optically thick lines, the drop in $\epsilon$ at high $n_{\rm H2}$ results from our assumption of fixed $\tau$. The optical depth for a given transition depends on the column density of molecules in the lower state for that transition ($J=0$ for most of our molecules, $J=1$ for CS). At densities near the peak, the molecules do not obey LTE. Sub-thermal excitation leads to a higher fraction of molecules in the lower, absorbing state. At higher densities, as LTE sets in, the column needed to reach our adopted $\tau$ will be higher because more molecules will be in the higher $J$ states. If we remake Figure \ref{fig:emis_vs_dens} fixing $N_{\rm mol}/\Delta v$ instead of $\tau$, then we do not see a similar drop in $\epsilon$ at high $n_{\rm H2}$ for optically thick lines. For optically thin lines, $\epsilon$ will still peak near the critical density for fixed $N_{\rm mol}/\Delta v$ because at higher $n_{\rm H2}$, near LTE, some additional potentially emitting molecules will be excited out of the emitting state.

In addition to collider density, optical depth sets $\epsilon$ at fixed $T_{\rm kin}$. At high $\tau$, emitted photons are re-absorbed rather than escaping, allowing more time for collisional de-excitation. This line trapping \citep{SCOVILLE74,GOLDREICH74} lowers the effective critical density of a line, with the well-known result that at high $\tau$, the effective critical density is adjusted by a factor $\tau^{-1}$ relative to the critical density in the absence of line trapping \citep[for recent treatments see][]{SHIRLEY15,SCOVILLE15}. 

Thus high $\tau$ lowers the $n_{\rm H2}$ for peak $\epsilon$ at a given $T_{\rm kin}$. The left panel of Figure \ref{fig:neff_vs_tau} illustrates the effect of changing $\tau$ on the normalized emissivity vs. $n_{\rm H2}$ plot seen in Figure \ref{fig:emis_vs_dens}. The right panel shows how changing $\tau$ affects the peak of the $\epsilon$ vs. density curve over the range where $\tau \approx 1$. For convenience, we include a machine readable table (Table \ref{tab:neff}) reporting the collider density, $n_{\rm H2}$, at which the emissivity reaches $95\%$ of its peak value given some $T_{\rm kin}$ and $\tau$.

At fixed $T_{\rm kin}$, the absolute value of $\epsilon$ depends on both $n_{\rm H2}$ and $\tau$. Figure \ref{fig:emis_twod} shows the interplay of these two factors for CO and HCN. We plot $\epsilon$, normalized to the peak of the whole plane, over a wide range of $\tau$ and $n_{\rm H2}$. The figure shows the decrease in $n_{\rm H2}$ for maximum $\epsilon$ with increasing $\tau$ (the peaks move left as one moves up). It also illustrates that at or around the critical density, $\epsilon$ becomes higher for lower $\tau$. As one would expect, the most emissive gas has low optical depth and $n_{\rm H2}$ near the critical density of the transition.

\begin{deluxetable}{lccc}
\tabletypesize{\scriptsize}
\tablecaption{Density for Maximum Emissivity in One Zone Models \label{tab:neff}}
\tablewidth{0pt}
\tablehead{
\colhead{Line} & 
\colhead{$T_{\rm kin}$} &
\colhead{$\log_{10} \tau$} &
\colhead{$\log_{10} n_{\rm eff}^{0.95}$} \\
\colhead{} & 
\colhead{[K]} &
\colhead{} &
\colhead{[cm$^{-3}$]}}
\startdata
12CO10 &  10.0 & -1.00 &  3.38 \\
12CO10 &  10.0 & -0.90 &  3.37 \\
12CO10 &  10.0 & -0.80 &  3.36 \\
12CO10 &  10.0 & -0.70 &  3.35 \\
12CO10 &  10.0 & -0.60 &  3.33 \\
12CO10 &  10.0 & -0.50 &  3.30 \\
12CO10 &  10.0 & -0.40 &  3.28 \\
12CO10 &  10.0 & -0.30 &  3.25 \\
12CO10 &  10.0 & -0.20 &  3.21 \\
12CO10 &  10.0 & -0.10 &  3.17 \\
\nodata & \nodata & \nodata & \nodata \\
\enddata
\tablecomments{The full version of this table is available as online only material. The table reports the minimum collider density at which the emissivity, $\epsilon$, reaches 95\% of its peak value for a given $T_{\rm kin}$ and $\tau$.}
\end{deluxetable}

\subsection{Emission for Density Distributions}
\label{sec:dist_results}

\begin{figure*}
\plotone{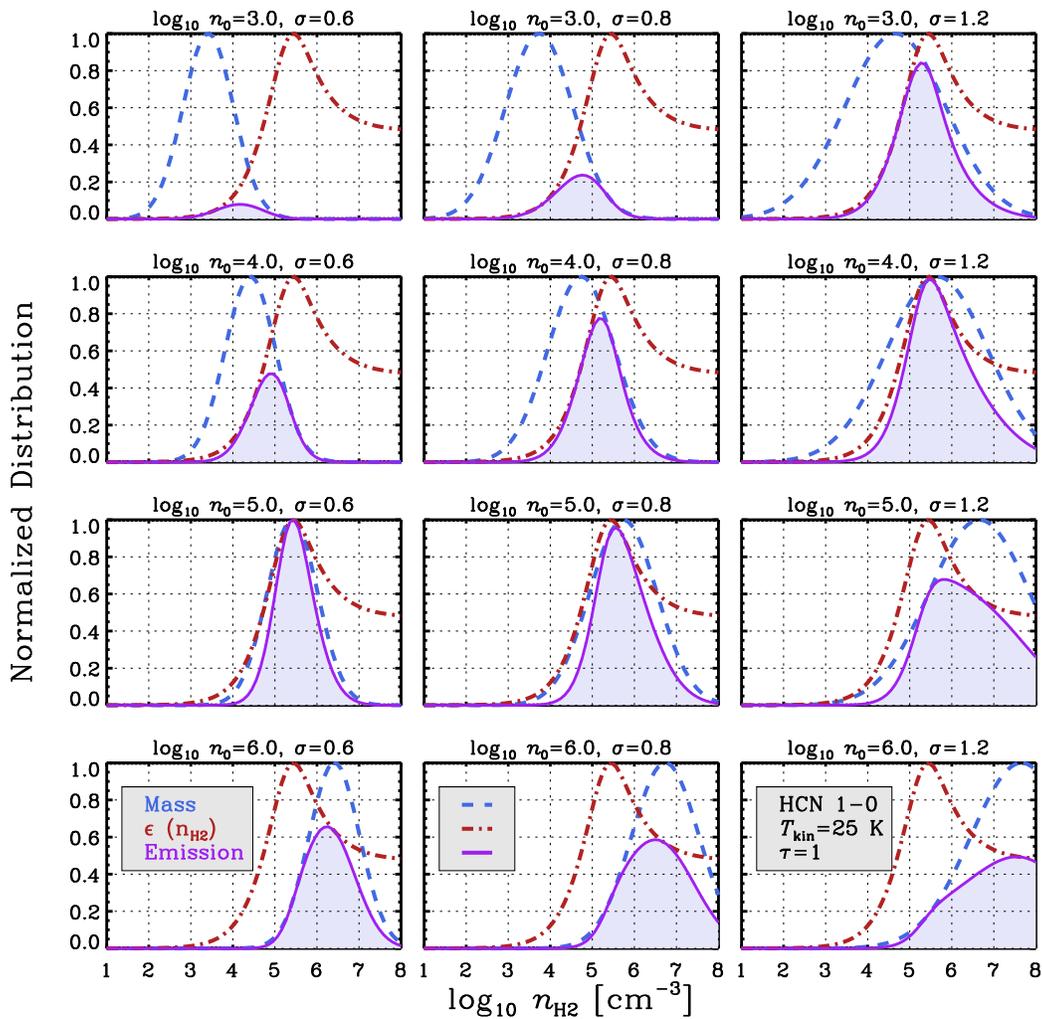}
\caption{Illustration of our model for HCN emission. The purple curves show emission as a function of collider density, $n_{\rm H2}$, for lognormal distributions with a range of mean densities, $n_0$ (in cm$^{-3}$), and widths ($\sigma$ in dex). The blue curves show the distribution of mass, the red curve --- which is the same in all panels --- shows $\epsilon (n_{\rm H2})$ for HCN at $T_{\rm kin} = 25$~K, $\tau = 1$. Their product, the purple curve, shows the light emitted as a function of density. Even as $\epsilon (n_{\rm H2})$ remains fixed, the density of gas emitting HCN~(1-0) changes dramatically with the sub-beam density distribution.}
\label{fig:pdf_illus}
\end{figure*}

We combine our one zone calculations to simulate emission from a distribution of densities within the beam. As described in \S \ref{sec:model}, we calculate the distribution of mass as a function of volume density, $n_{\rm H2} P(n_{\rm H2})$ for realistic distributions of densities. Then we multiply the mass distribution by the one zone emissivity shown in the previous section. From this product, we calculate the beam-averaged emissivity, \beameps , and the density below which $50\%$ of the light is emitted for each spectral line, $n_{\rm med}^{\rm emis}$ (see \S \ref{sec:defs}).

Figure \ref{fig:pdf_illus} illustrates the interplay of emissivity and density distribution at the core of the model. The red curve shows $\epsilon (n_{\rm H2})$ of HCN for $T_{\rm kin} = 25$~K and $\tau = 1$. The blue curves show a series of lognormal distributions with varying mean density, $n_0$, and width, $\sigma$. The product of the emissivity and the mass distribution, shown as a purple curve, indicates how much emission emerges from each density. The plots take $\epsilon$ and the mass distribution to be normalized at their peak values, but we do not normalize the emission, so that the purple curves can be compared among panels to see how the integrated brightness of the line varies.

\subsection{What Densities Produce Emission?}
\label{sec:med_dens}

\begin{figure*}
\plottwo{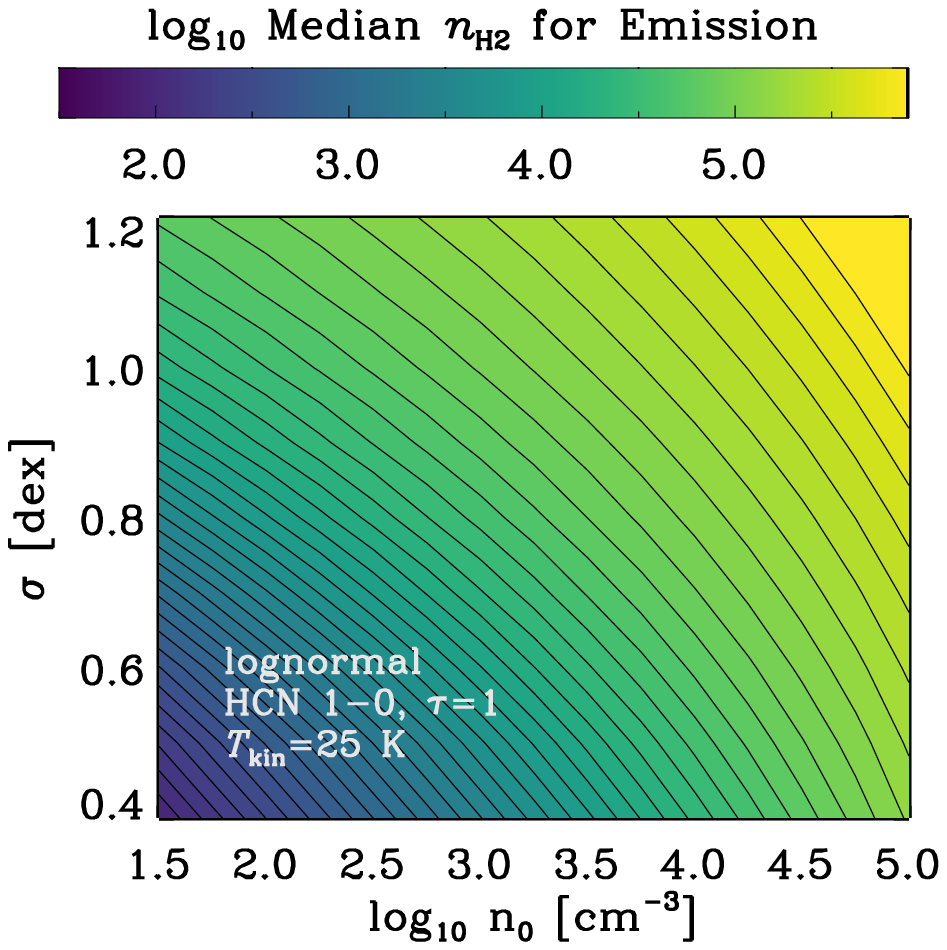}{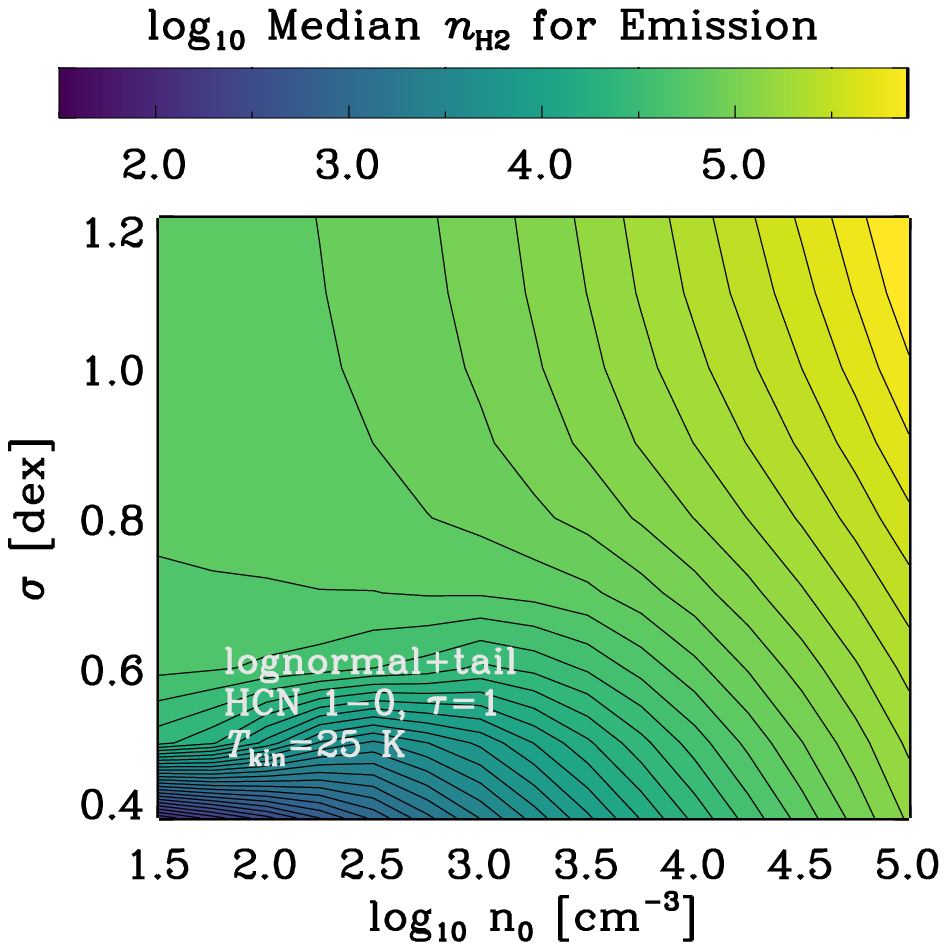}
\caption{Median density for emission of HCN~(1-0), $n_{\rm med}^{\rm emis}$, as a function of the mean density, $n_0$, and width, $\sigma$, for lognormal density distributions without ({\em left}) and with ({\em right}) a power law tail (\S \ref{sec:distdef}). Contours are spaced by $0.1$~dex. The range of $\sigma$ corresponds to turbulent Mach numbers $\mathcal{M} \approx 2.5{-}100$. For pure lognormal distributions, the median density for HCN emission varies strongly with the combination of $n_0$ and $\sigma$. It can reach low values, $n_{\rm H2} \lesssim 10^{3}{-}10^4$~cm$^{-3}$, for values of $n_0$ and $\sigma$ observed for extragalactic molecular clouds. A power law tail, if present, changes the picture ({\rm right} panel). Such a tail ensures the presence of some high density, high emissivity material even for low $n_0$. In this case, $n_{\rm med}^{\rm emis}$ remains $\gtrsim 10^{4.5}$~cm$^{-3}$ except for very narrow, low $n_0$ distributions. Because the power law tail now sets the high density mass distribution, the influence of $\sigma$ also decreases when such a tail is present.}
\label{fig:neff_dist}
\end{figure*}

\begin{figure*}
\plottwo{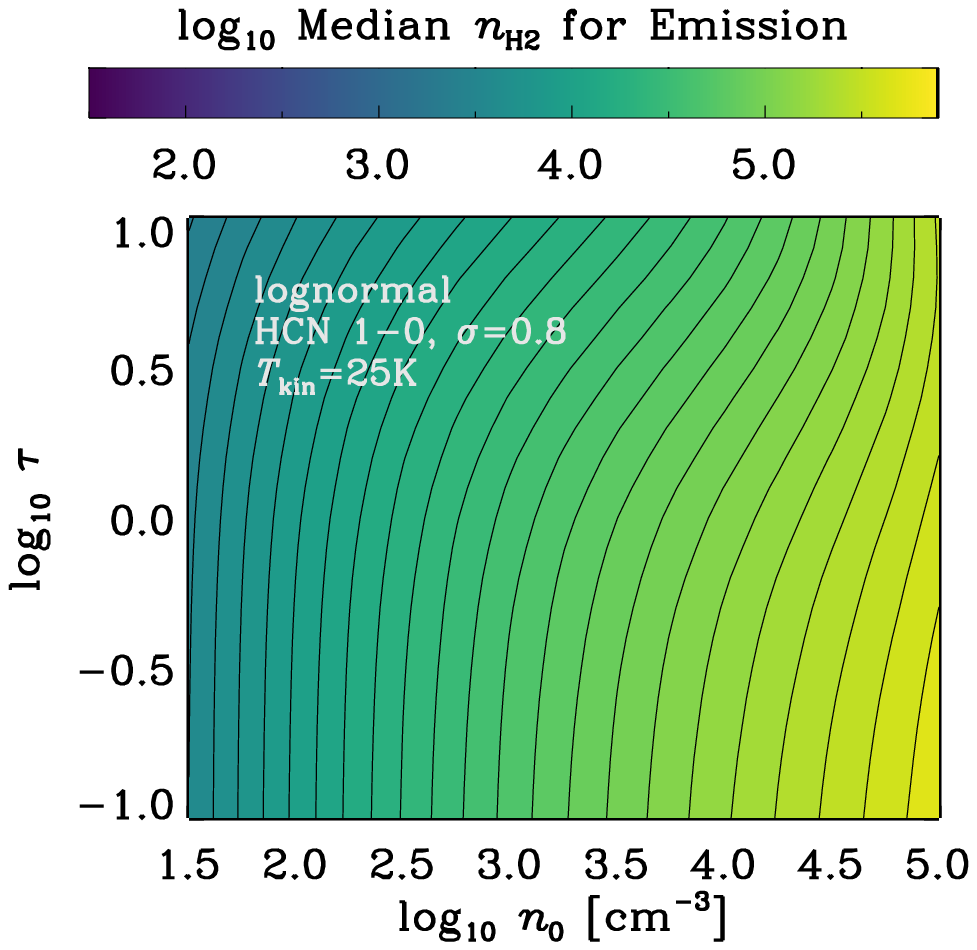}{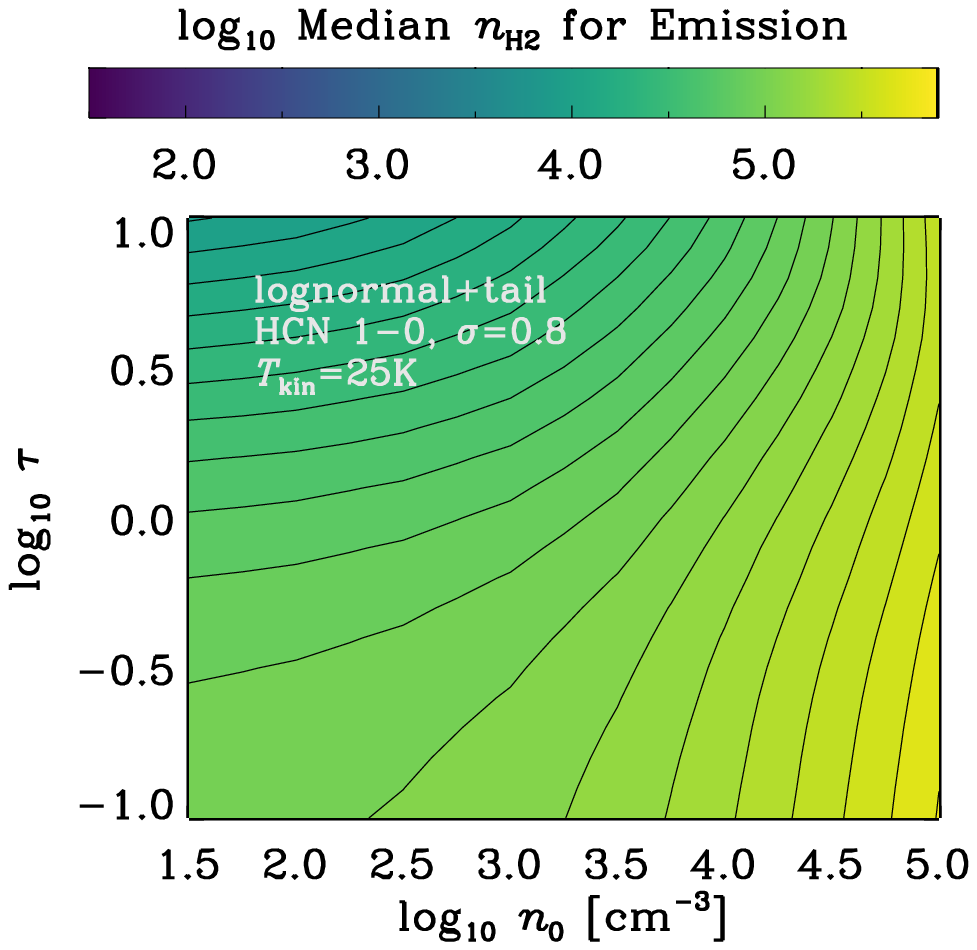}
\caption{As Figure \ref{fig:neff_dist}, but now plotting optical depth, $\tau$, on the $y$-axis while holding the distribution width fixed at $\sigma=0.8$~dex. Optical depth plays a modest role for the pure lognormal case ({\rm left} panel) at low $n_0$. It has a stronger impact on $n_{\rm med}^{\rm emis}$ when the density distribution includes a power law tail ({\rm right} panel). In this case a high optical depth allows gas with densities on the tail to emit effectively at lower $n_{\rm H2}$ and so lowers $n_{\rm med}^{\rm emis}$.}
\label{fig:neff_tau}
\end{figure*}

Figure \ref{fig:pdf_illus} illustrates how the density of gas producing HCN emission depends on the sub-beam density distribution. Perhaps most strikingly, when the gas at densities far below the critical density significantly outmasses gas at high density (as in the top left panels), most HCN emission comes from modest density gas. When much of the mass lies at high $n_{\rm H2}$, as in the bottom row, the density for emission tracks the density of the gas and HCN traces the total gas reservoir, as CO does for lower density clouds \citep[see][]{KRUMHOLZ07B,NARAYANAN08}.

We identify the characteristic density for emission, $n_{\rm med}^{\rm emis}$, as the density below which half the emission emerges. For HCN, we plot $n_{\rm med}^{\rm emis}$ as a function of $n_0$, $\sigma$, and $\tau$ in Figures \ref{fig:neff_dist} and \ref{fig:neff_tau}. Figure \ref{fig:neff_dist} shows the effect of varying the mean density, $n_0$, and distribution width, $\sigma$ on $n_{\rm med}^{\rm emis}$ for models with and without a power law tail. The range of $\sigma$ spans from $0.4{-}1.2$, corresponding to $\mathcal{M} \approx 2.5{-}100$. Although we illustrate the case for HCN, other dense gas tracers show similar behaviors with shifted density scales.

For a lognormal distribution (left panel in Figure \ref{fig:neff_dist}), the median density for HCN emission depends on both the mean, $n_0$, and the width, $\sigma$. A high $n_0$ shifts the distribution of masses to higher $n_{\rm H2}$. A high $\sigma$ does the same, both by shifting the peak of the mass distribution and by extending the wing of the distribution to higher densities. Both effects yield higher $n_{\rm med}^{\rm emis}$ (see \S \ref{sec:distdef} or compare across a row in Figure \ref{fig:pdf_illus}), and for pure lognormal distributions, $\sigma$ and $n_0$ exert similar effects. This interplay leads to the curved contours in the left panel of Figure \ref{fig:neff_dist}. To achieve a given $n_{\rm med}^{\rm emis}$ a distribution can either have high $\sigma$ or high $n_0$ or both.

The absolute value of $n_{\rm med}^{\rm emis}$ varies over a wide range in the left panel of Figure \ref{fig:neff_dist}. Lognormal distributions with low $n_0$ and low $\sigma$ will generate most of their HCN emission from low $n_{\rm med}^{\rm emis} \lesssim 10^3{-}10^4$~cm$^{-3}$. Such distributions do not have high beam averaged emissivity (the purple curve is low in Figure \ref{fig:pdf_illus}), but they could be common. Cloud-averaged mean densities $\lesssim 10^{2.5}$~cm$^{-3}$ and Mach numbers $\mathcal{M} \lesssim 5{-}10$ are widespread in nearby galaxies \citep[e.g.,][]{BOLATTO08,HUGHES13B,LEROY16}. If the sub-beam density distributions in such clouds were pure lognormals, then much of their HCN emission could arise from gas at moderate densities, in contrast to the normal extragalactic view of HCN as a high density tracer.

The right panel of Figure \ref{fig:neff_dist} shows that the inclusion of a power tail in the density distribution (see \S \ref{sec:distdef}) changes this picture somewhat. Such a tail can add high $n_{\rm H2}$, high $\epsilon$ material to clouds with otherwise low mean densities. When we include such a tail in the calculation, the range of $n_{\rm med}^{\rm emis}$ narrows considerably across the parameter space in Figure \ref{fig:neff_dist}. As long as $\sigma \gtrsim 0.6$~dex ($\mathcal{M}\gtrsim 5$), the median density for emission is in the range $n_{\rm H2} \gtrsim 10^{4.5}$~cm$^{-3}$. For narrower density distributions ($\sigma \lesssim 0.6$~dex) with modest $n_0$, our formulation of the power law tail does not add as much high density material and $n_{\rm med}^{\rm emis}$ still reaches low values. This may only reflect an issue with our adopted threshold for the onset of the power law; we fix this at a multiple of $n_0$, while a more general physical prescription might reflect $\sigma$ and other physical properties of the gas \citep[e.g., see][]{KRUMHOLZ05,FEDERRATH12}.

The inclusion of a power law tail diminishes the influence of $\sigma$ on $n_{\rm med}^{\rm emis}$ compared to the pure lognormal case. Though this results partially from our model formulation, the physical point is general: in the case where self-gravitating structures dominate the high end of the density distribution, the turbulent Mach number may have diminished effect on emission from high density tracers. This can be tested, for example, by correlation of density sensitive ratios like HCN/CO or CS/CO with measurements of the turbulent velocity dispersion, e.g., from high resolution CO imaging.

Figure \ref{fig:neff_dist} adopts a fixed optical depth, $\tau = 1$, for HCN. Variations in $\tau$ will also affect $n_{\rm med}^{\rm emis}$. In Figure \ref{fig:neff_tau} we plot $n_{\rm med}^{\rm emis}$ for HCN, holding $\sigma$ fixed at $0.8$~dex but allowing $\tau$ to vary from $0.1$ to $10$. Again the left panel shows results for a pure lognormal, and the right shows results for a distribution with a power law tail at high densities. The overall sense of both panels is that a high $\tau$ leads to a lower $n_{\rm med}^{\rm emis}$, but that the effects are modest over the plausible range of $\tau$ for HCN.

For the lognormal case, $\tau$ exerts a weaker influence than variations in $\sigma$. Variations across the plausible range of $\tau$ (optically thin up to $\tau \sim 10$) change $n_{\rm med}^{\rm emis}$ by only a factor of a few at fixed $n_0$. Optical depth has a stronger influence when a power law tail is present and the mean density, $n_0$, is low. In this case, a higher $\tau$ shifts the density for maximum emission (the effective critical density) to lower $n_{\rm emis}$. This allows lower density material on the tail to emit effectively. The result is that $\tau$ plays a main role in setting $n_{\rm med}^{\rm emis}$ for modest $n_0 \lesssim 10^3$~cm$^{-3}$ when a power law tail is present.

The plots here illustrate $n_{\rm med}^{\rm emis}$ for HCN. Similar trends hold for our other lines. In general, the interplay of a realistic density distribution and $\epsilon (n_{\rm H2})$ can lead to a wide range of emitting densities for lognormal distributions. Including a power law tail, however, mitigates the variation for many cases. Optical depth exerts a more modest influence on $n_{\rm med}^{\rm emis}$ than the density distribution in the lognormal case, but can play a significant role when a power law tail is present and $n_0$ is modest. 

We include tabulated results for $n_{\rm med}^{\rm emis}$ as a function of $\tau$, $n_0$, $\sigma$, and $T_{\rm kin}$ for each line in Table \ref{tab:dist_tab}. These can be used to reproduce the results of this and the following sections.

\begin{deluxetable*}{lcccccccccc}
\tabletypesize{\scriptsize}
\tablecaption{Beam Averaged Emissivity and Density Distributions \label{tab:dist_tab}}
\tablewidth{0pt}
\tablehead{
\colhead{Line} & 
\colhead{Distribution} & 
\colhead{$\log_{10} n_0$} &
\colhead{$\sigma$} &
\colhead{$\log_{10} X_{\rm mol}$} &
\colhead{$\log_{10} n_{\rm med}^{\rm mass}$} &
\colhead{$\log_{10} f_{\rm dense}$} &
\colhead{$T_{\rm kin}$} &
\colhead{$\log_{10} \tau$} &
\colhead{$\log_{10} \left< \epsilon \right>$} &
\colhead{$\log_{10} n_{\rm med}^{\rm emis}$} \\
\colhead{} & 
\colhead{} & 
\colhead{[cm$^{-3}$]} &
\colhead{[dex]} &
\colhead{} &
\colhead{[cm$^{-3}$]} &
\colhead{} &
\colhead{[K]} &
\colhead{} &
\colhead{[$\frac{{\rm K~km~s}^{-1}}{{\rm cm}^{-2}}$]} &
\colhead{[cm$^{-3}$]} \\
\colhead{(1)} & 
\colhead{(2)} & 
\colhead{(3)} &
\colhead{(4)} &
\colhead{(5)} &
\colhead{(6)} &
\colhead{(7)} &
\colhead{(8)} &
\colhead{(9)} &
\colhead{(10)} &
\colhead{(11)} \\
}
\startdata
12CO10 & LOGNORMAL+TAIL &  2.00 &   0.6 & -4.00 &  2.41 & -1.63 &  15.0 & -1.00 & -19.20 &  2.78 \\
12CO10 & LOGNORMAL+TAIL &  2.00 &   0.7 & -4.00 &  2.60 & -1.31 &  15.0 & -1.00 & -19.14 &  2.98 \\
12CO10 & LOGNORMAL+TAIL &  2.00 &   0.8 & -4.00 &  2.80 & -1.14 &  15.0 & -1.00 & -19.10 &  3.15 \\
12CO10 & LOGNORMAL+TAIL &  2.00 &   0.9 & -4.00 &  2.98 & -1.03 &  15.0 & -1.00 & -19.07 &  3.27 \\
12CO10 & LOGNORMAL+TAIL &  2.00 &   1.0 & -4.00 &  3.11 & -0.96 &  15.0 & -1.00 & -19.06 &  3.36 \\
12CO10 & LOGNORMAL+TAIL &  2.00 &   1.1 & -4.00 &  3.22 & -0.92 &  15.0 & -1.00 & -19.04 &  3.42 \\
12CO10 & LOGNORMAL+TAIL &  2.00 &   1.2 & -4.00 &  3.30 & -0.88 &  15.0 & -1.00 & -19.04 &  3.47 \\
12CO10 & LOGNORMAL+TAIL &  2.25 &   0.6 & -4.00 &  2.66 & -1.50 &  15.0 & -1.00 & -19.11 &  2.93 \\
12CO10 & LOGNORMAL+TAIL &  2.25 &   0.7 & -4.00 &  2.85 & -1.19 &  15.0 & -1.00 & -19.08 &  3.12 \\
12CO10 & LOGNORMAL+TAIL &  2.25 &   0.8 & -4.00 &  3.04 & -1.01 &  15.0 & -1.00 & -19.05 &  3.28 \\
\nodata & \nodata & \nodata & \nodata & \nodata & 
\nodata & \nodata & \nodata & \nodata & \nodata \\
\enddata
\tablecomments{The full version of this table is available as online only material. The table reports results for line emission for distributions of densities observed within a beam. Columns report: (1) emission line, (2) distribution shape, (3) mean density for lognormal part of density distribution, (4) rms dispersion in lognormal part of density distribution, (5) adopted abundance of species relative to H$_2$, (6) median density by mass for the adopted distribution, (7) dense gas mass fraction, defined as gas above $n_{\rm H2} = 10^{4.5}$~cm$^{-3}$, (8) adopted kinetic temperature, (9) adopted optical depth, (10) calculated beam-averaged emissivity of H$_2$, (11) median density for emission.}
\end{deluxetable*}

\subsection{Beam-Averaged Emissivity and Line Ratio Patterns}
\label{sec:ratios}

\begin{figure*}
\plottwo{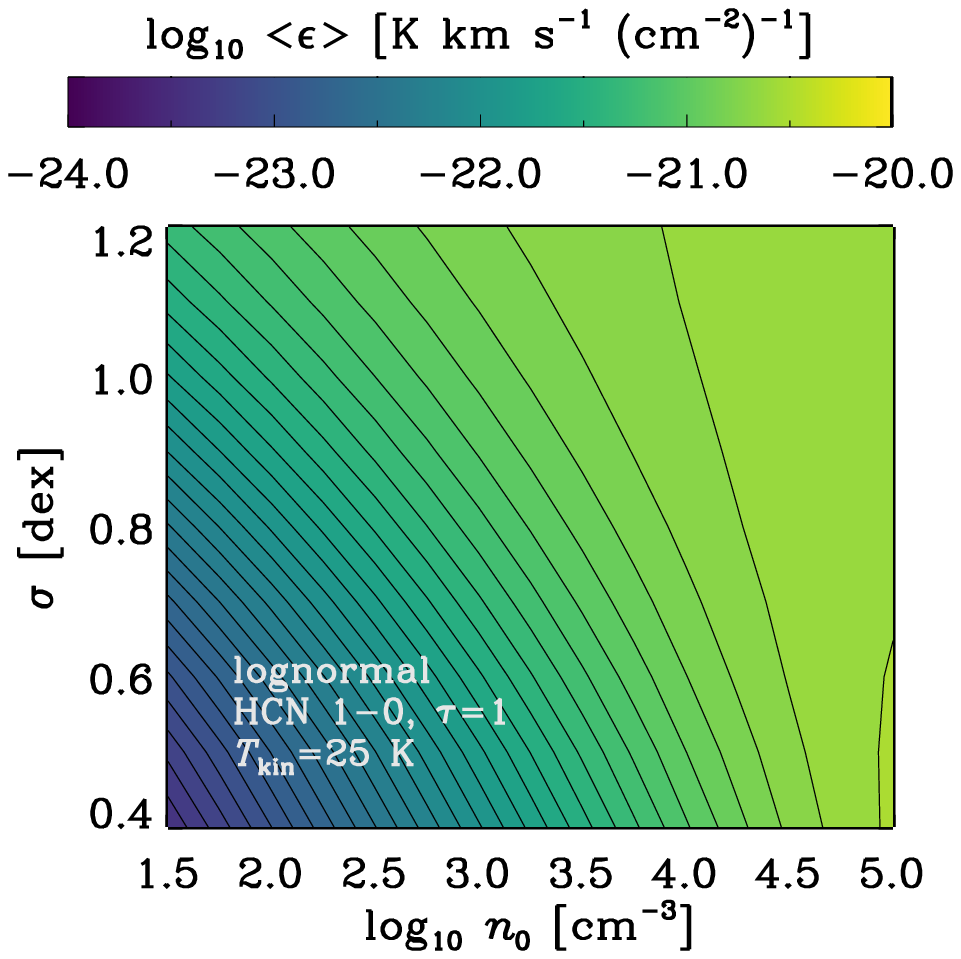}{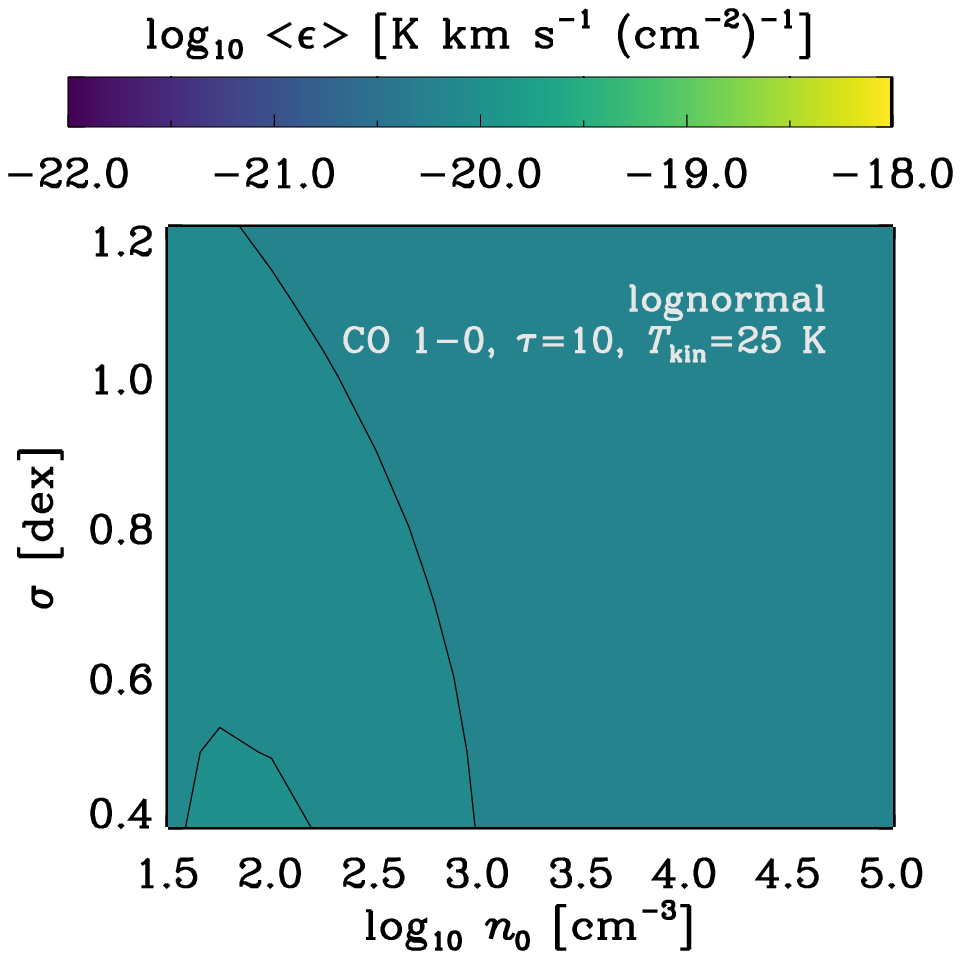}
\caption{Beam-averaged emissivity, \beameps , of HCN~(1-0) ({\em left}) and CO~(1-0) ({\em right}) for z lognormal density distribution as a function of the mean, $n_0$, and width, $\sigma$, of the distribution. As expected, the emissivity of HCN varies according to the fraction of high density material, which is higher for high $n_0$ and high $\sigma$. Optically thick CO exhibits only weak emissivity variations, consistent with its use as a tracer of the total H$_2$ column. The ratio of the two lines provides a diagnostic of the mass of high density material in the beam.}
\label{fig:beameps}
\end{figure*}

The convolution of emissivity and the density distribution determines the overall ability of the gas to emit. For each model, we record the beam-averaged emissivity, \beameps , defined in Equation \ref{eq:beamemis}. \beameps\ measures emission per unit mass averaged over the whole density distribution, and so quantifies how well a given density distribution emits in a given transition. In Figure \ref{fig:pdf_illus}, \beameps\ corresponds to the integral of the purple curves, and so varies across the figure.

Figure \ref{fig:beameps} shows \beameps\ as a function of $n_0$ and $\sigma$ for HCN~(1-0) and $^{12}$CO~(1-0) at their fiducial $\tau$. For HCN~(1-0) shown in the left panel, \beameps\ changes by more than two orders of magnitude across the plotted parameter space. The sense of the variation follows Figure \ref{fig:pdf_illus}: high $n_0$ and high $\sigma$ lead to higher \beameps\ because they yield more dense gas in the distribution. The right panel shows \beameps\ for CO~(1-0) at $\tau=10$. In contrast to HCN~(1-0), CO~(1-0) shows little variation in \beameps , reflecting that the low densities needed to excite emission are present across the whole model grid. These weak variations in \beameps\ help motivate the use of low $J$ CO as a tracer of the total H$_2$ column.

\begin{figure*}
\plottwo{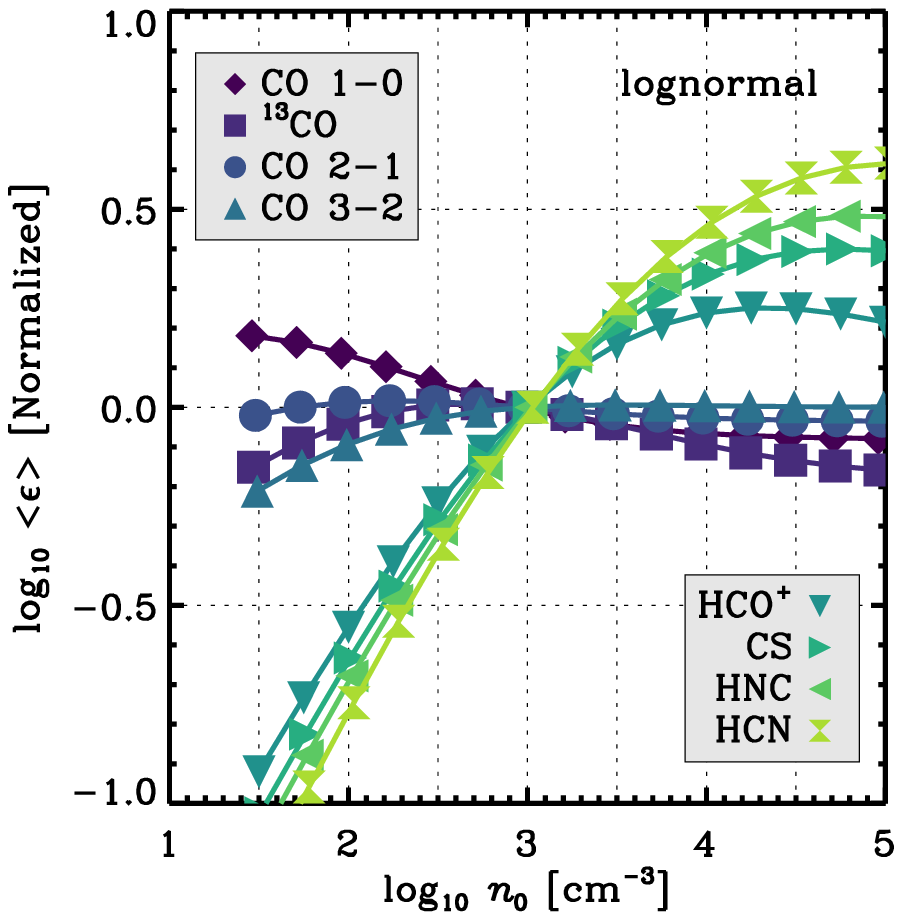}{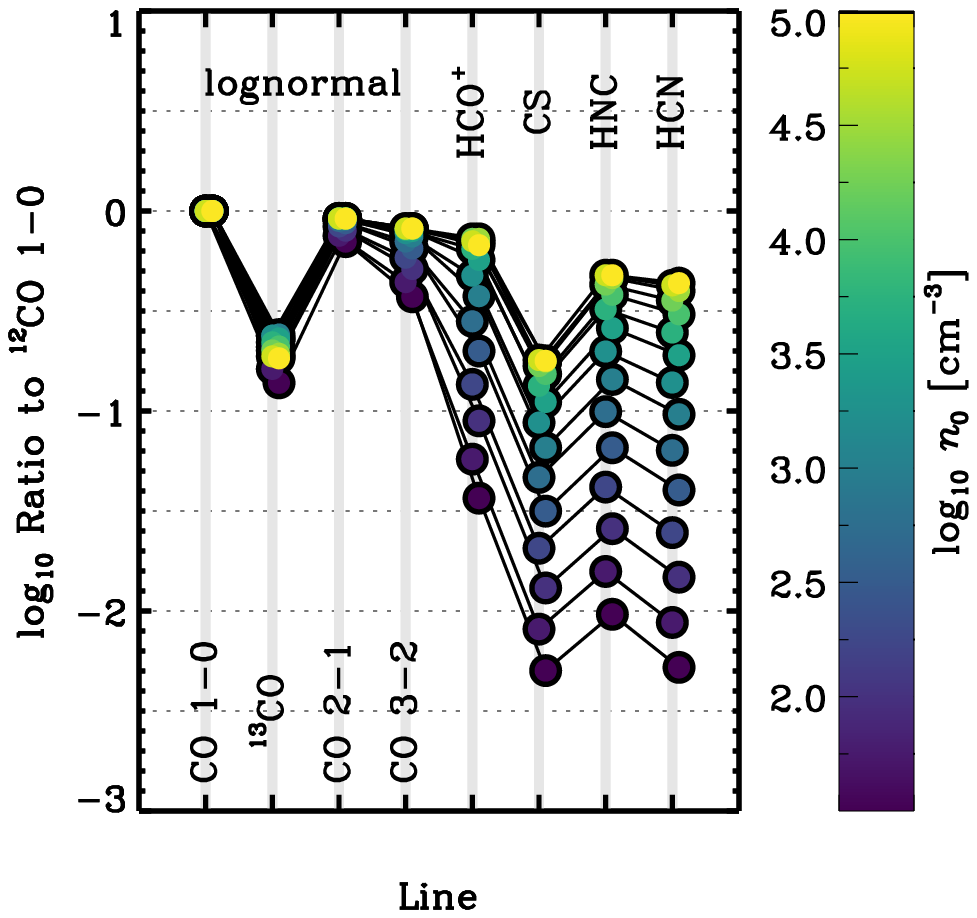}
\plottwo{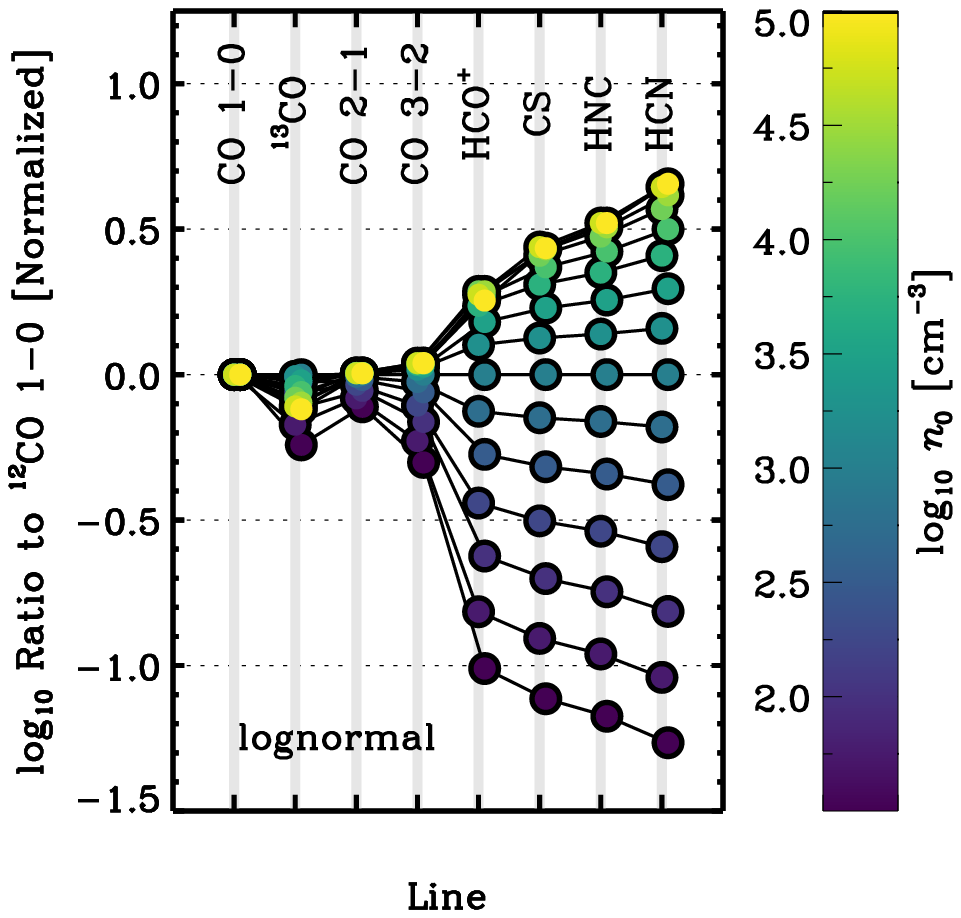}{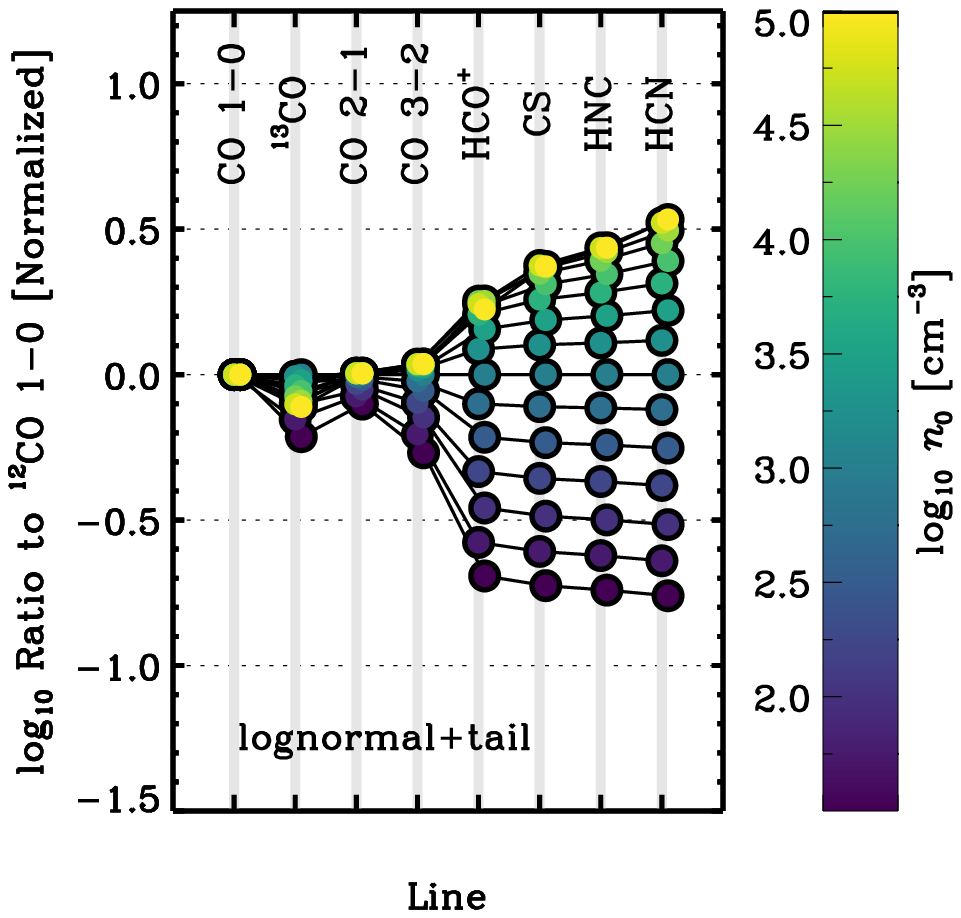}
\caption{Beam averaged emissivity, \beameps , and line ratio patterns as a function of changing mean density, $n_0$. We adopt our fiducial $\tau$ and take $\sigma = 0.8$~dex and $T_{\rm kin} = 25$~K throughout. ({\em top left}) \beameps\ as a function of $n_0$ for each line, normalized to \beameps\ for that line at $n_0 = 10^3$~cm$^{-3}$. Here we plot results for pure lognormal distributions. The emissivity of a distribution in the CO lines varies only weakly with its mean density. However, \beameps\ for the high density tracers depends strongly on the sub-grid density distribution. ({\em top right}) Implied line ratio pattern relative to CO for our fiducial abundances and different $n_0$, again for pure lognormal distributions. Line ratios between dense gas tracers and CO lines vary strongly with density. Internal ratios among CO lines and dense gas tracers vary more weakly. The exact pattern depends on our adopted abundances. ({\em bottom row}) Line ratio patterns, now normalized to the pattern at $n_0 = 10^3$~cm$^{-3}$, for various $n_0$. We calculate results for both a pure lognormal distribution ({\em bottom left}) and one with a power law tail ({\em bottom right}). These variations in line ratios are robust to the absolute abundance pattern, but will be affected by abundance variations. For both distributions, the most density-sensitive lines show the strongest variations (e.g., HCN \mbox{(1-0)}, HNC \mbox{(1-0)}, CS \mbox{(2-1)}) relative to CO. In the lognormal case, a changing density distribution creates a ``flaring'' pattern of line ratio variations when the lines are sorted by effective critical density. This signature is far less pronounced when the power law tail dominates the density distribution (e.g., this is the case at low $n_0$ in the bottom right plot).}
\label{fig:linerat}
\end{figure*}

\begin{figure}
\plotone{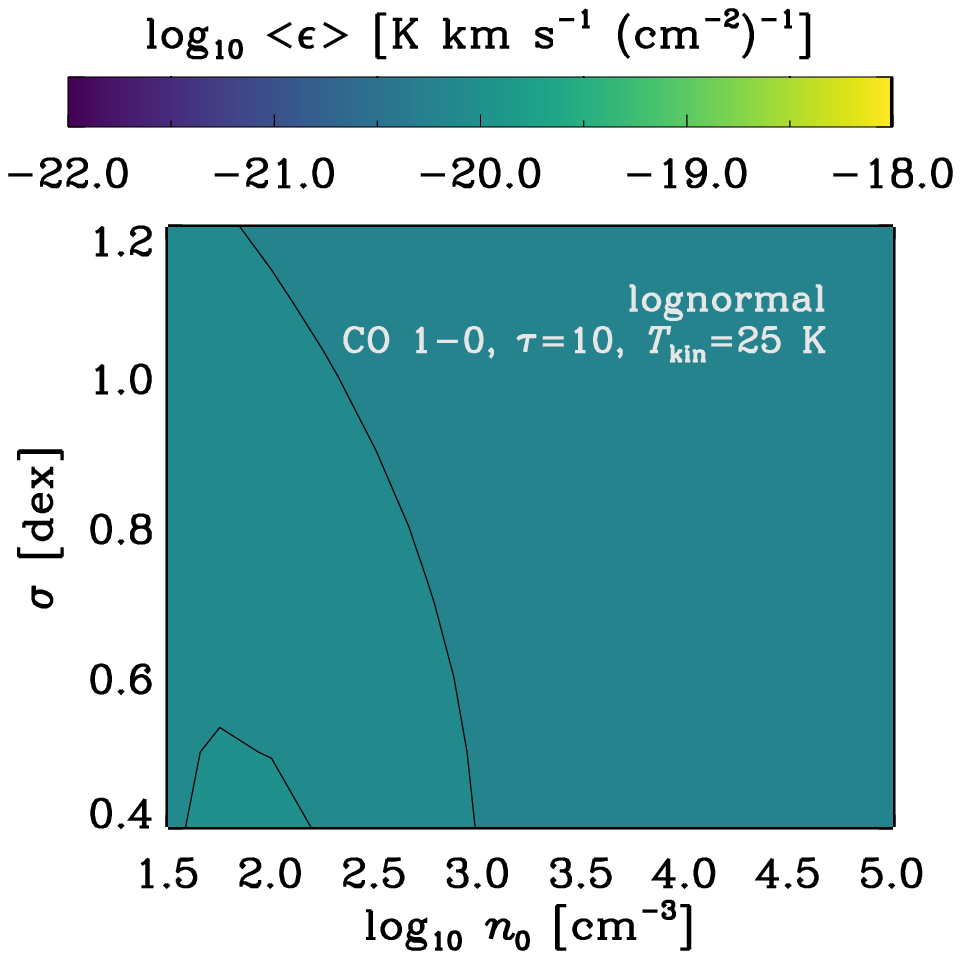}
\caption{As the bottom row of Figure \ref{fig:linerat} but now varying $T_{\rm kin}$ while other quantities remain fixed. We plot line ratios relative to those expected for $T_{\rm kin} = 35$~K, so that the figure shows the expected changes to the observed line ratio pattern due to temperature variations. These are much weaker than the density variations seen in Figure \ref{fig:linerat}.}
\label{fig:linerat_temp}
\end{figure}

The top left panel in Figure \ref{fig:linerat} shows how $\beameps$ changes with mean density, $n_0$, for each of our target lines. This plot assumes a pure lognormal distribution with $\sigma = 0.8$~dex ($\mathcal{M} \sim 11$). The density-sensitive transitions (HCN \mbox{(1-0)}, HNC \mbox{(1-0)}, HCO$^+$ \mbox{(1-0)}, CS \mbox{(2-1)}), show strong variations in \beameps\ as $n_0$ changes. These transitions are much brighter in denser gas. As a result, \beameps\ increases by more than an order of magnitude as $n_0$ goes from $\sim 10^2$~cm$^{-3}$ to $10^5$~cm$^{-3}$. Meanwhile, the CO transitions show weak variations in \beameps\ across the same range of $n_0$.

By observing ratios between the dense gas tracers and the CO lines, one can leverage these differences in \beameps\ to constrain the sub-beam density distribution. The top right panel of Figure \ref{fig:linerat} shows \beameps\ for each line divided by \beameps\ for CO~(1-0). These \beameps\ ratios can be directly observed as line ratios within a matched beam. That is, the top right panel of Figure \ref{fig:linerat} shows predicted line ratio patterns for lognormal density distributions with a range of mean densities, $n_0$.

The pattern in Figure \ref{fig:linerat} changes with $n_0$. Line ratios between high density tracers and CO~(1-0) change by more than an order of magnitude as $n_0$ varies. Variations among transitions that trace similar densities are much smaller. For example, ratios among the CO lines change by a factor of $\lesssim 2$ as $n_0$ changes more than two orders of magnitude.

The exact ratios in Figure \ref{fig:linerat} depend on our adopted abundances, which are often poorly known. Even if we do not know the true abundances, they may remain approximately fixed among targets. In this case, comparing line ratios offers a way to diagnose changes in physical conditions. That is, translating an HCN/CO ratio to a dense gas fraction requires knowledge of the abundance of HCN and CO. But comparing two HCN/CO ratios to one another may allow a comparative statement about the mean density or the dense gas fraction in which factors like abundance divide out.

The bottom panels in Figure \ref{fig:linerat} illustrate this approach for a pure lognormal ({\em left}) and a lognormal distribution with a power law tail ({\em right}). Again we plot ratios relative to CO~(1-0). Now, however, we normalize each line ratio pattern by the pattern calculated for $n_0 = 10^3$~cm$^{-3}$. Thus, the figure shows how we expect line ratios to change as $n_0$ changes. For example, if $n_0$ changes from $10^3$~cm$^{-3}$ to $10^4$~cm$^{-3}$, then we expect HCN/CO to increase by $0.5$~dex, about a factor of $3$. These changes are independent of the absolute value of the adopted abundances, they require only that the abundance remains fixed between locations.

In the next section, we explore the quantitative relationship between changing line ratios and changes in the dense gas fraction or median density. Before proceeding, we note some qualitative behavior from Figure \ref{fig:linerat}. First, large changes in $n_0$ induce large changes in the ratios between dense gas tracers and total gas tracers. This is the strongest behavior in the figures. Internal ratios among CO lines or dense gas tracers show relatively weak contrasts, but a change in $n_0$ leads to significant changes between any dense gas tracer and CO.

Second, the magnitude of this variation depends in detail on the sub-grid density distribution adopted. That is, the pure lognormal distributions ({\em left}) and those that include a power law tail at high densities ({\em right}) show distinct results. The model with a power law tail shows weaker variations in line ratios across the same range of $n_0$. It also shows less variation among the different high density tracers, especially at low mean density (the bluer colors). For the lower end of $n_0$, $\sim 10^2 {-} 10^3$ cm$^{-3}$, the four high density tracers all show about the same strength of variations despite their different effective critical densities.

Meanwhile, the pure lognormal shows strong variations in line ratios that ``flare'' as a function of effective critical density. That is, variations in line ratio as a function of $n_0$ are strongest for our highest critical density line, HCN. They appear weakest for HCO$^+$, which has the lowest effective critical density among our dense gas tracers. This flaring pattern is also evident in the lognormal with a power law, but the behavior is much weaker and appears more prominently at high $n_0$ than at low $n_0$, where it is almost absent.

The differences between the pure lognormal and the case with a power law tail can be understood from the shapes of the two distributions. To first order, each tracer picks up material at or above some effective critical density. In the power law case, as long as two densities both lie on the power law tail, the ratio of material above these two densities will be fixed. That is:

\begin{equation}
\label{eq:massratio}
\frac{M (n > n_1)}{M (n > n_2)} = \frac{\int_{n_1}^{\infty} n^{\alpha} dn}{\int_{n_2}^{\infty} n^{\alpha} dn} \propto \frac{n_1^{1+\alpha}}{n_2^{1+\alpha}}~.
\end{equation}

\noindent Where we have used the definitions above\footnote{Recall that $\alpha$ is the slope in logarithmic units, so that the integral in normal space goes as $\int n \times n^{\alpha-1} \propto \int n^{\alpha}$.} and take $\alpha < -1$. Equation \ref{eq:massratio} illustrates that once $\alpha$ and two threshold densities on the power law tail are determined, then the ratio of mass above these two densities on the power law tail will be fixed. In other words, the power law itself is scale free: $n_0$ only affects the lognormal part and the interface between the two.

No such fixed ratio holds for the lognormal. The integral above some threshold density falls rapidly with increasing threshold ($\propto 0.5 ( 1- {\rm erf}~n)$). As a result, for any pair of densities, the ratio will change as the mean of the distribution changes. The sense of the change is that the mass above a threshold density close to (but still above) the mean density changes less than the mass above a threshold density at a high value. Thus, for the lognormal distribution the lines with highest effective critical densities (HCN, HNC) show the strongest variations in their ratios. Those with lower effective critical densities show weaker variations. This leads to the flaring seen in Figure \ref{fig:linerat}. It will also lead to a non-linear mapping between any given line ratio and dense gas fraction, assuming that dense gas fraction is defined relative to some fixed threshold density.

Temperature variations create weaker changes in the line ratio pattern than density variations. Figure \ref{fig:linerat_temp} shows the effect of varying T$_{\rm kin}$ while other inputs to the model remain fixed. Over the range $T_{\rm kin} = 10{-}50$~K, line ratio variations induced by changing temperature show lower magnitude and less flaring than those that we calculate for changing $n_0$. This reflects that a higher $T_{\rm kin}$ also renders the CO line brighter. A higher $T_{\rm kin}$ does have some effect on the line ratio pattern, though, because high $T_{\rm kin}$ lowers the critical density and so makes it easier to excite emission from dense gas tracer at intermediate densities.

{\em Implication for Observations:} Figures \ref{fig:linerat} and \ref{fig:linerat_temp} show how one can approach observations of an ensemble of lines \citep[e.g., as in the EMPIRE survey;][]{BIGIEL16}. Density variations should induce large changes in ratios between the CO lines and high density tracers. When the lines are sorted by effective critical density, changes in the mean density for a lognormal or similar steep and curving distribution should produce flaring line ratio variations. That is, the lines with highest effective critical density respond most strongly to density changes. When a power law tail is present, such flaring will be weaker, and the lines that sample the power law tail should change in lock step. The line ratio variations themselves will also be weaker in the presence of a power law tail, because such a tail ensures the presence of some high density gas.

Thus, ensembles of lines can in principle be used to constrain the characteristic shape of the sub-resolution density distribution. Another application is to use a suite of lines to help control for abundance variations. In the bottom row of Figure \ref{fig:linerat}, we have assumed the abundance of each molecule to remain fixed between beams, though we have not assumed any particular abundance values. In the case that one line exhibits a distinct behavior while the others indicate density variations, abundance variations (which do happen) represent a logical hypothesis.

Figure \ref{fig:linerat} does illustrate an issue with our target line suite. Our lines sort into two groups, which broadly exhibit similar behaviors. All four CO lines show weak line ratio changes with density, while all four dense gas tracers show strong variations. We have chosen the $\lambda \approx 3$~mm lines most readily observable in other galaxies, but this suite leaves us with limited sensitivity to intermediate densities.

\subsection{Quantitative Estimates of Changing Gas Density}
\label{sec:deltas}

\begin{deluxetable}{lcccc}
\tabletypesize{\scriptsize}
\tablecaption{Line Ratio Predictors of Gas Density \label{tab:predict}}
\tablewidth{0pt}
\tablehead{
\colhead{} &
\multicolumn{2}{c}{Lognormal + Tail} & 
\multicolumn{2}{c}{Pure Lognormal} \\
\colhead{Line Relative} & 
\colhead{Slope\tablenotemark{a}} &
\colhead{Scatter\tablenotemark{b}} &
\colhead{Slope\tablenotemark{a}} &
\colhead{Scatter\tablenotemark{b}} \\
\colhead{to CO~(1-0)} & 
\colhead{} &
\colhead{(dex)} &
\colhead{} &
\colhead{(dex)} 
}
\startdata
\hfill\\[-10pt]
\cutinhead{Predicting $f_{\rm dense}$}
HCO$^+$& $1.21$ & $0.19$ & $2.31$ & $0.35$ \\
CS & $1.08$ & $0.18$ & $1.86$ & $0.33$ \\
HNC & $1.05$ & $0.15$ & $1.70$ & $0.27$ \\
HCN & $0.98$ & $0.16$ & $1.41$ & $0.26$ \\
\cutinhead{Predicting median $n_{\rm H2}$ by mass}
HCO$^+$& $2.03$ & $0.44$ & $1.68$ & $0.51$\\
CS & $1.85$ & $0.41$ & $1.54$ & $0.47$ \\
HNC & $1.83$ & $0.35$ & $1.47$ & $0.42$\\
HCN & $1.73$ & $0.36$ & $1.34$ & $0.42$
\enddata
\tablecomments{Power law index relating the change in line ratio to the change in median H$_2$ density by mass, $n_{\rm med}^{\rm mass}$, and dense gas fraction, $f_{\rm dense}$. Models run over mean densities from $\log_{10} n_0 [{\rm cm}^{-3}] = 2$ to $4$, $T_{\rm kin} = 15$ to $35$~K, and $\sigma = 0.6$ to $1.2$~dex. We report results with and without a power law tail separately. Models assume our fiducial optical depths.}
\tablenotetext{a}{Best fit slope, $\alpha$, of a scaling $y = \alpha x$, that goes through the origin, with $x$ the logarithmic change in line ratios, e.g., $\log_{10}$~(HCN/CO)$_1 - \log_{10}$~(HCN/CO)$_2$, and $y$ the logarithmic change in $n_{\rm med}^{\rm mass}$.}
\tablenotetext{b}{Indicative logarithmic scatter about the power law scaling. Reflective of ability of ratio to predict changes in gas density. For comparison, the whole model grid (with no fit) has scatter ${\sim} 1$~dex in $n_{\rm med}^{\rm mass}$ and ${\sim} 0.5{-}0.7$~dex in $f_{\rm dense}$.}
\tablenotetext{c}{Fraction of mass at densities above $n_{\rm H2} = 10^{4.5}$~cm$^{-3}$.}
\end{deluxetable}

Beyond the qualitative analysis in the previous section, we would like to use mm-wave line ratios to infer quantitative changes in the underlying density distribution. We expect changes in the ratio, e.g., of HCN-to-CO to reflect differences in the fraction of dense gas, $f_{\rm dense}$. Assuming that some characteristic density distribution holds from beam-to-beam, these ratios will also reflect changes in the median density of gas by mass, $n_{\rm med}^{\rm mass}$. 

We test our ability to infer changes in $f_{\rm dense}$ and $n_{\rm med}^{\rm mass}$ using a grid of models. Across this grid, we vary the mean density from $n_0 = 10^2$~cm$^{-3}$ to $10^4$~cm$^{-3}$, the distribution width from $\sigma = 0.6$ to $1.2$~dex ($\mathcal{M} \approx 5$ to $100$), and the temperature from $T_{\rm kin} = 15$ to $35$~K. These values of $n_0$ and $\sigma$ span most of the values observed for whole clouds in normal and starburst galaxies \citep[e.g.,][]{LEROY15A,LEROY16}. We calculate line ratio patterns for each model in this grid (Table \ref{tab:dist_tab}). Then for all possible model pairs, we measure the logarithmic change in each line ratio, 

\begin{equation}
\Delta \log_{10} {\rm HCN/CO} = \frac{\log_{10} I_{\rm HCN,1}/I_{\rm CO,1}}{\log_{10} I_{\rm HCN,2} / I_{\rm CO,2}}~,
\end{equation}

\noindent where the subscripts 1 and 2 refer to some pair of models in the grid and $\Delta \log_{10} {\rm HCN/CO}$ is expressed in dex.

For each model pair, we also measure the logarithmic change in the median density by mass, $n_{\rm med}^{\rm mass}$, and the dense gas mass fraction. Here, $n_{\rm med}^{\rm mass}$ is defined as the density below which half the mass in the distribution lies. The dense gas fraction, $f_{\rm dense}$, is defined as the fraction of mass above some threshold density, $n_{\rm thresh}$. We take $n_{\rm thresh} = 10^{4.5}$~cm$^{-3}$ as our default threshold and explore the appropriate definition of $n_{\rm thresh}$ below.

\subsubsection{Estimating the Dense Gas Fraction}

\begin{figure*}
\plottwo{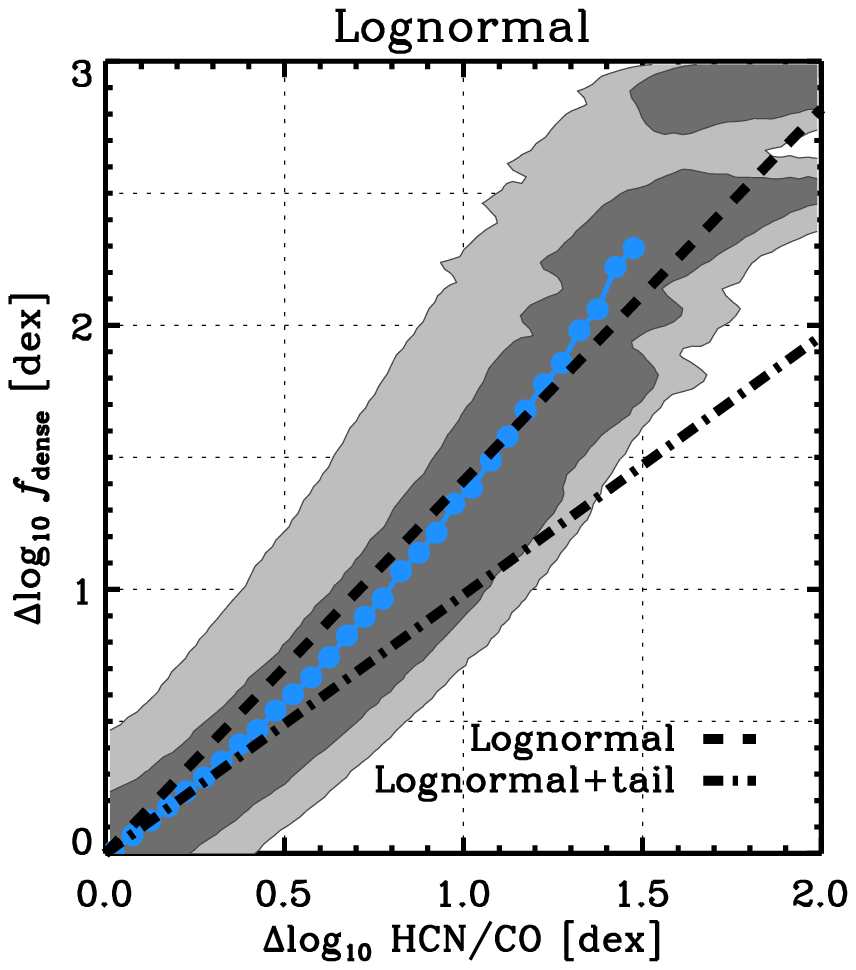}{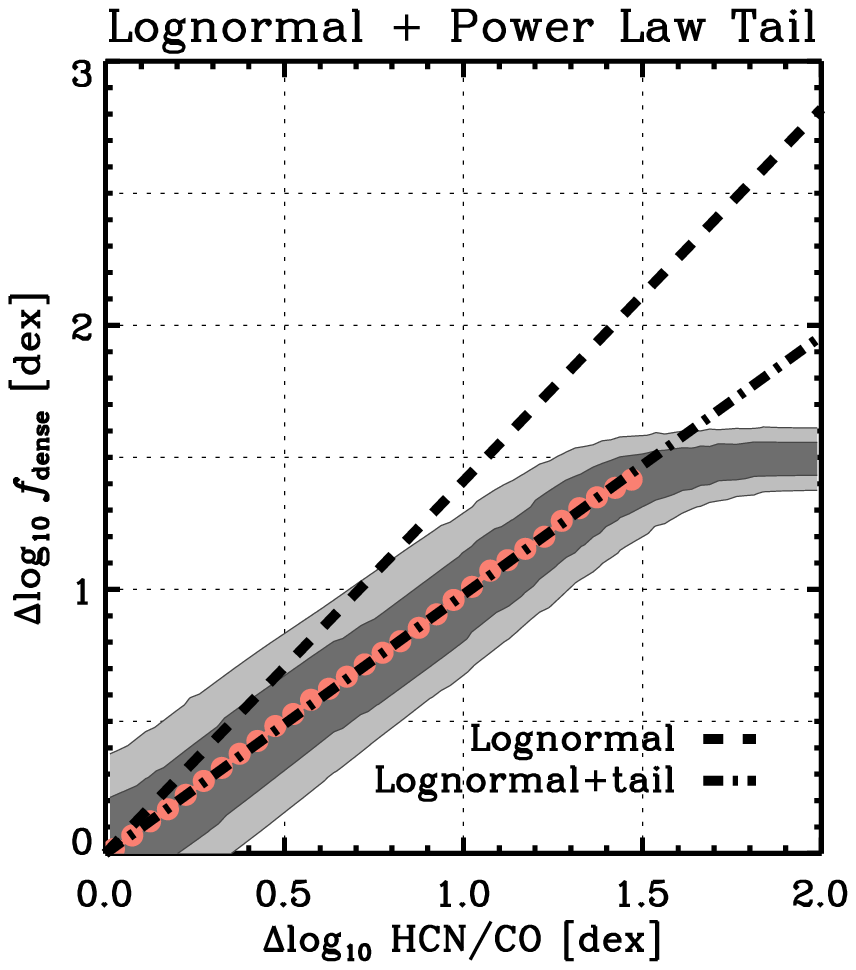}
\caption{Mapping between changes in HCN-to-CO line ratio ($x$-axis) to changes in the dense gas mass fraction $f_{\rm dense}$ (defined relative to $n_{\rm thresh} > 10^{4.5}$~cm$^{-3}$). We calculate cloud-integrated emissivities for a range of densities ($\log_{10} n_0 = 2{-}4$~cm$^{-3}$), distribution widths ($\sigma = 0.6{-}1.2$ dex), and temperatures ($T_{\rm kin} = 15{-}35$~K). Then, we pair calculations and measure the logarithmic contrast in line ratios and dense gas fractions. Contours show the $68\%$ and $95\%$ range of model pairs at for each change in line ratio. Colored points show binned results. Lines in both panels indicate the best fit scaling for $\Delta \log_{10} {\rm HCN/CO} < 1.5$}
\label{fig:lrat_fdense}
\end{figure*}

\begin{figure*}
\plottwo{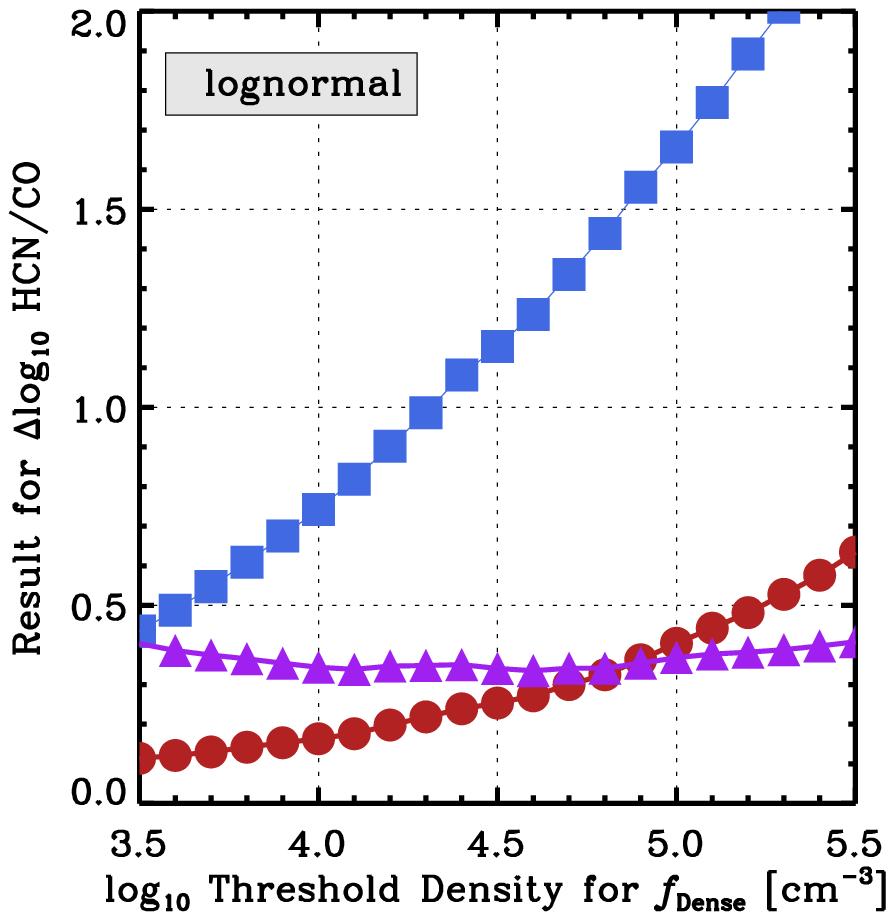}{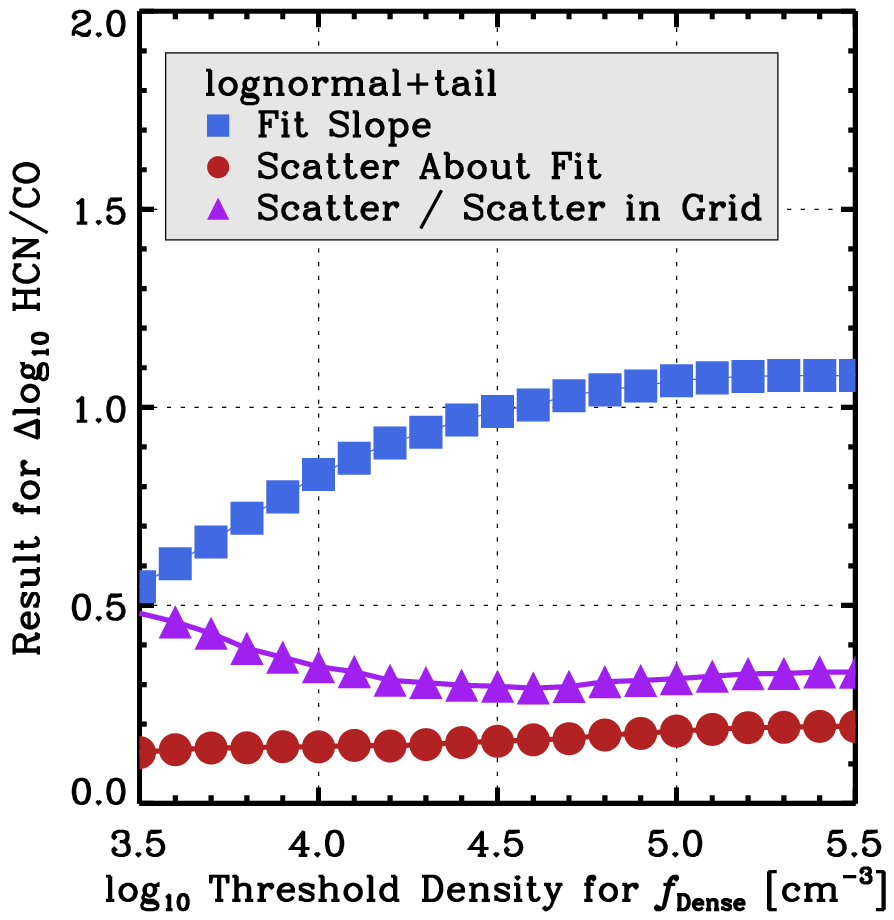}
\caption{Effect of choice of density threshold to define $f_{\rm dense}$ on the ability of HCN/CO to predict $f_{\rm dense}$. Best fit slope, scatter, and improvement over scatter in the model grid for various definitions of $f_{\rm dense}$ for a pure lognormal ({\em left}) and a lognormal with a power law tail ({\em right}). When a power law tail is present, our results are robust to choice of threshold above $\sim 10^{4.5}$~cm$^{-3}$, while the appropriate fit for the pure lognormal depends sensitively on the adopted threshold.}
\label{fig:pick_thresh}
\end{figure*}

A main use for ratios like HCN/CO is to estimate the fraction of gas mass that is dense. Figure \ref{fig:lrat_fdense} shows how well $\Delta \log_{10} {\rm HCN/CO}$ predicts changes in the dense gas mass fraction across our model grid. We plot results for pure lognormal distributions ({\em left}) and distributions with power law tails ({\em right}) separately. Figure \ref{fig:lrat_fdense} adopts a threshold density $n_{\rm thresh} = 10^{4.5}$~cm$^{-3}$.

In both cases, $\Delta \log_{10} {\rm HCN/CO}$ can predict changes in $f_{\rm dense}$, but the two distributions yield different results. For the case with a power law tail $\Delta \log_{10} {\rm HCN/CO}$ relates to $f_{\rm dense}$ via a roughly linear relation. We expect such behavior based on the arguments in \S \ref{sec:deltas}, with the ratio in the mass above any two densities on the power law fixed for fixed $\alpha$. For most of our models, both HCN and $n_{\rm thresh}$ used to define $f_{\rm dense}$ sample power law part of the distribution. In this case we expect a simple, linear relation between the dense gas tracer and $f_{\rm dense}$. The scatter in our model reflects temperature variations and cases where the lognormal part of the distribution contributes to the calculation.

The relationship between $f_{\rm dense}$ and $\Delta \log_{10} {\rm HCN/CO}$ appears steeper for the pure lognormal ({\em left} panel of Figure~\ref{fig:lrat_fdense}). Following our argument in \S \ref{sec:deltas}, the steep, curving form of the lognormal renders the ratio of area above any two densities sensitive to the mean of the distribution. This, in turn, leads to a steeper than linear relation between $f_{\rm dense}$ and $\Delta \log_{10} {\rm HCN/CO}$.

We plot results for HCN/CO, but report results for all of our dense gas tracers in Table \ref{tab:predict}. Following the logic above, for the power law tail case, all tracers show slopes near unity with modest ($\approx 50\%$) scatter across the model grid. When a fixed-slope power law tail is present, dense gas tracers behave as expected and do a good job of tracing the fraction of mass in dense gas. Tracers with lower effective critical densities show mildly non-linear slopes and shallow scatter as the lognormal portion of the distribution begins to contribute to the emission in some cases. For a pure lognormal, the relationships between $\Delta \log_{10} {\rm HCN/CO}$ and $\Delta \log_{10} f_{\rm dense}$ are steeper, with power law index from $1.4$ to $2.3$. In this case, the line ratio variations predict $f_{\rm dense}$ with larger scatter, a factor of $\sim 2$ across the grid.

Figure \ref{fig:lrat_fdense} shows results for dense gas defined by a threshold density $n_{\rm thresh} = 10^{4.5} \approx 3 \times 10^4$~cm$^{-3}$. This is also the definition adopted by \citet{GAO04B}. Figure \ref{fig:pick_thresh} shows how the results depend on our definition of dense gas. We vary $n_{\rm thresh}$ and calculate (1) the best fit power law slope relating $\Delta~f_{\rm dense}$ to $\Delta$~HCN/CO, (2) the scatter among all model pairs about this best-fit relation, and (3) the scatter in $\Delta \log_{10} f_{\rm dense}$ relative to the total scatter in the model grid. This latter is relevant because for low values of $n_{\rm thresh}$ the grid itself shows little variation; most models have a large fraction of their mass above $n_{\rm thresh} \approx 10^{3.5}$~cm$^{-3}$. The contrast between the scatter about the fit and the scatter in the grid highlight the accuracy with which HCN/CO picks out $f_{\rm dense}$.

Figure \ref{fig:pick_thresh} again shows distinct behavior for the pure lognormal ({\em left}) and the case with a power law tail at high density ({\em right}). In the case of a power law tail, any threshold density above $\log_{10} n_{\rm H2} \approx 4.5$~cm$^{-3}$ yields a nearly linear relation between $\Delta$ HCN/CO and $\Delta f_{\rm Dense}$. In the case of the pure lognormal, the exact scaling depends strongly on the threshold density chosen. In the pure lognormal case, high threshold densities yield a steep slope and larger scatter about the fit. 

Thus, when a fixed-slope power law tail is present in the density distribution, $\Delta \log_{10} {\rm HCN/CO}$ and other dense gas tracers appear to be stable tracers of the dense gas mass fraction for most reasonable definitions.  They capture $\Delta \log_{10} f_{\rm dense}$ with modest scatter over a range of plausible temperature, density, and Mach number variations. For the case of a pure lognormal distribution, the relationship between line ratios and $f_{\rm dense}$ is more unstable, and depends sensitively on adopted definitions. We return to these different results for different distributions below. Briefly, we prefer the case with a power law tail as more physical, given the presence of ongoing star formation in most regions targeted by extragalactic surveys. But addressing this question represents a natural next topic for observational and theoretical work.

\subsubsection{Estimating the Median Density}

\begin{figure*}
\plottwo{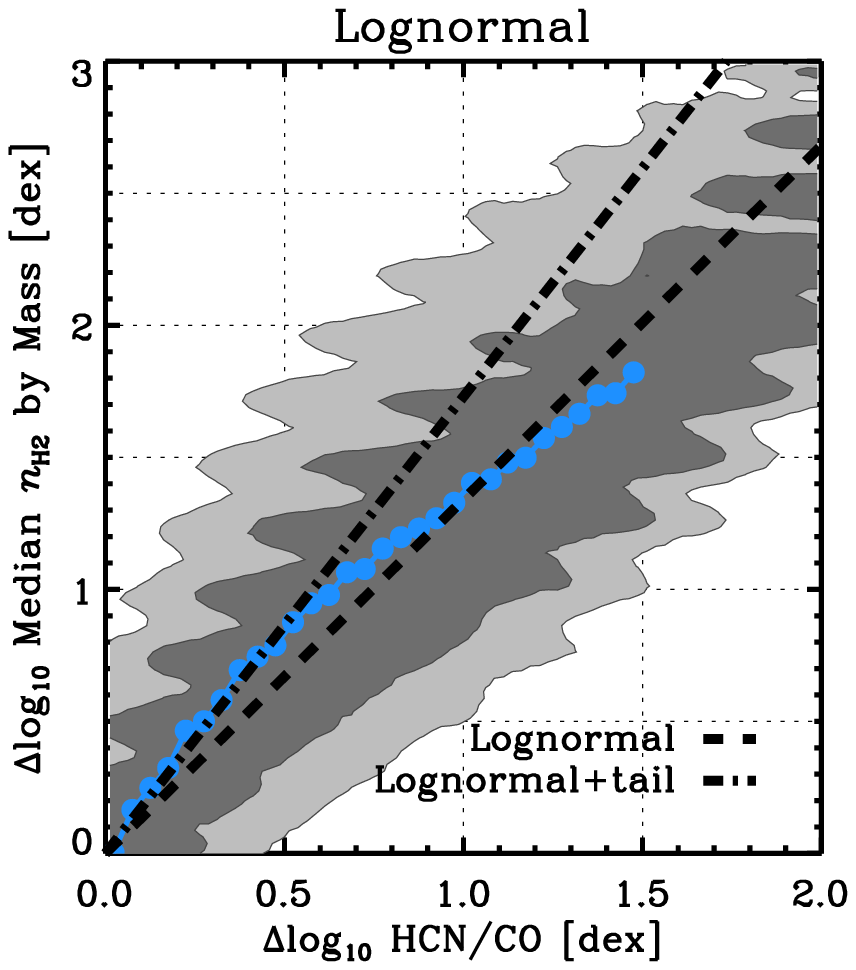}{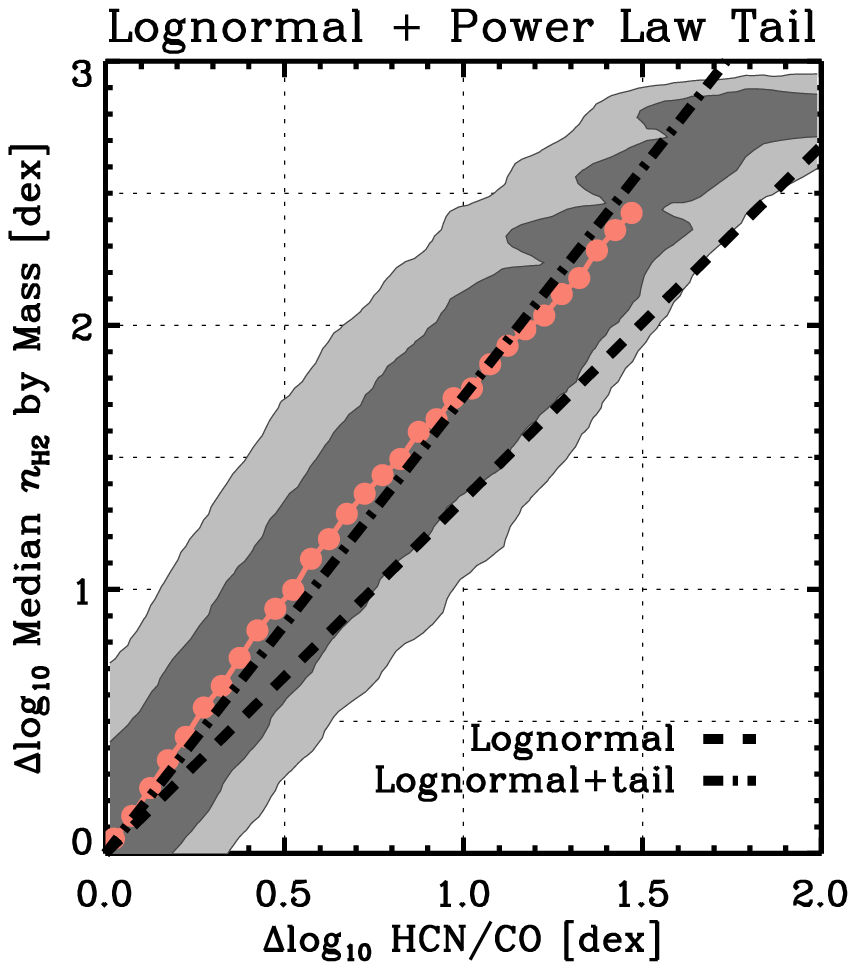}
\caption{Using changes in the HCN/CO line ratio to infer changes in the median $n_{\rm H2}$ by mass. As Figure \ref{fig:lrat_fdense}, but now both panels relate HCN-to-CO line ratio ($x$-axis) to changes in $n_{\rm med}^{\rm mass}$ ($y$-axis).}
\label{fig:lrat_nmed}
\end{figure*}

The CO lines constrain the integral under the distribution, while dense gas tracers access the integral only above some high density. If the shape of the density distribution is known, combining these two pieces of information constrains $n_{\rm med}^{\rm mass}$. 

Figure \ref{fig:lrat_nmed} shows results from our model grids using pairs of models to relate the logarithmic change in $n_{\rm med}^{\rm mass}$ to $\Delta \log_{10} {\rm HCN/CO}$. Again, gray contours enclose 68\% and 95\% of the model pairs at a given $\Delta \log_{10} {\rm HCN/CO}$. The colored points show the models binned by change in line ratio. Black lines show the best fit slope for change in line ratios $< 1.5$~dex. Table \ref{tab:predict} reports the best fit slopes and scatter for each dense gas tracer.

Changing line ratios can predict changes in the median $n_{\rm H2}$ by mass across our model grid. However, they do so with a factor of $\sim 2{-}3$ rms accuracy and the slope of the best fit scaling depends on the adopted sub-beam distribution. To first order, this calculation works by taking the integral above the effective critical density of some dense gas tracer, normalizing by the integral under the whole distribution (traced by CO), and relating this to the median of the distribution. The mapping is not perfectly one to one across the model grid, but does work.

This calculation works about equally well for the two distributions, but it seems better posed for the pure lognormal. In this case, the whole distribution of densities obeys a single simple functional form. Thus, it makes sense to constrain the whole distribution by contrasting the high density wing and the integral under the whole curve. In the case of the power law tail, the results rely sensitively on the adopted threshold joining the two distributions and slope of the power law tail.

\subsubsection{Using Ensembles of Lines}

\begin{figure*}
\plottwo{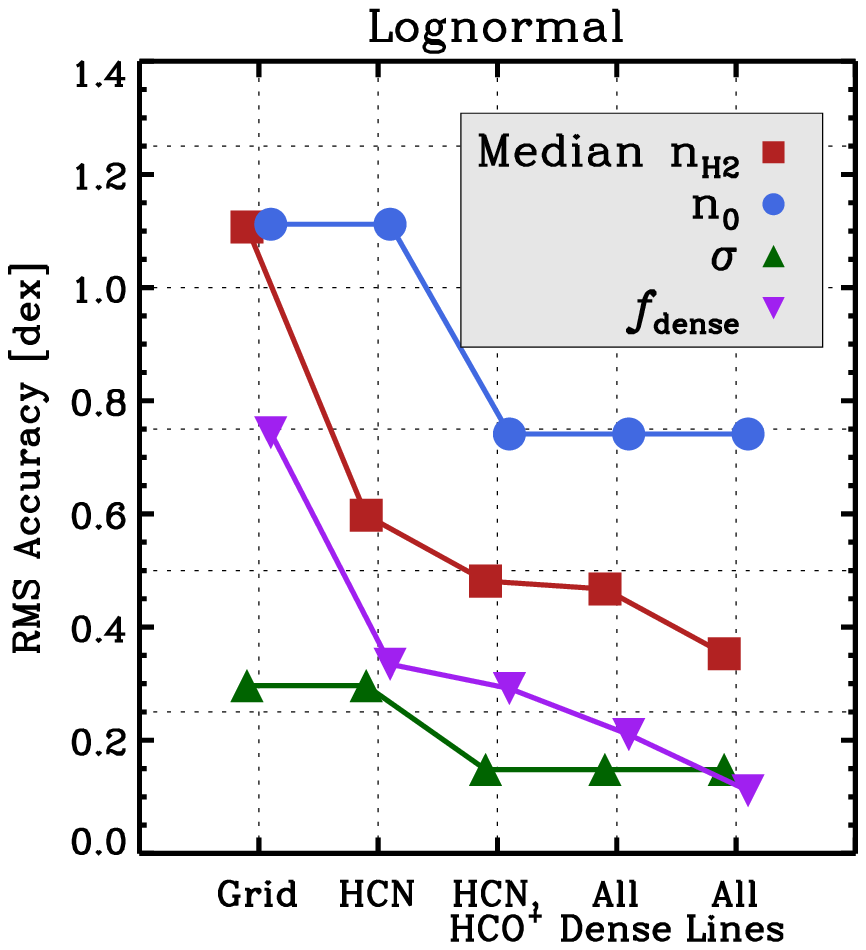}{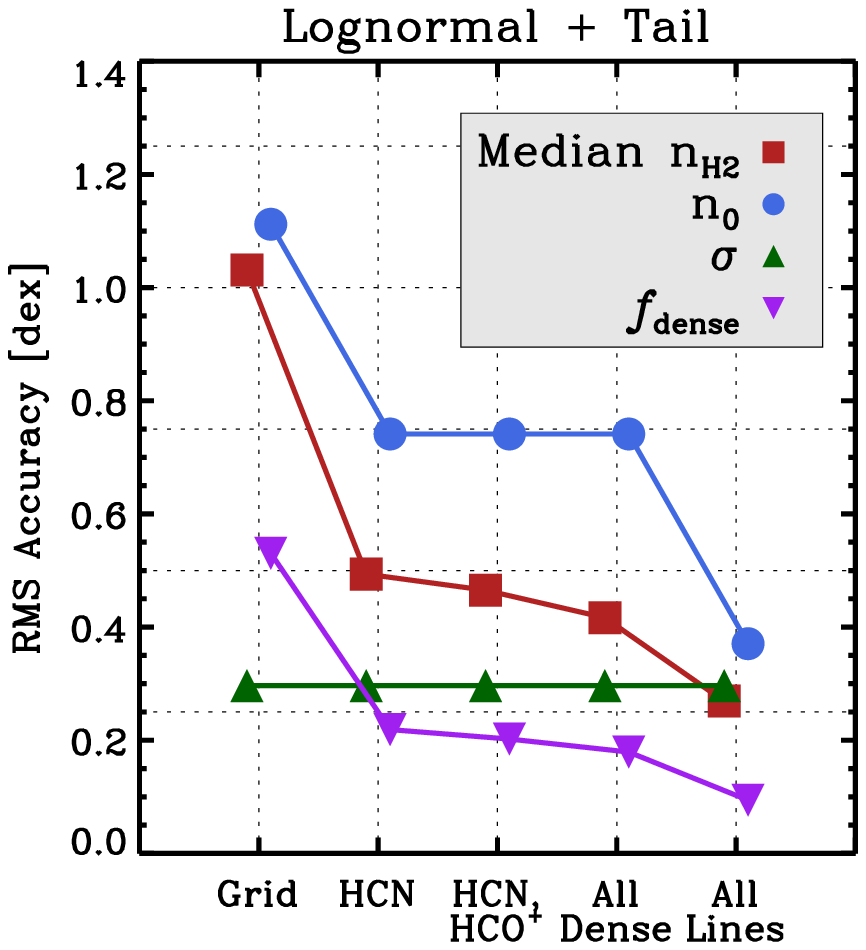}
\caption{Accuracy of recovered model inputs requiring all line ratios to match the model within a tolerance of $0.05$~dex. If fitting observations, these would also be good fits to the data. The two panels show results for a pure lognormal ({\em left}) and a lognormal distribution with a powerlaw tail ({\em right}). From left to right in each panel, we show the scatter in the model grid itself (not a fit), a fit using only the HCN/CO ratio, and then fits adding in HCO$^+$, all dense gas tracers, and all lines. The figure shows that $f_{\rm dense}$ can be recovered most accurately. The median density by mass, $n_{\rm med}^{\rm mass}$, can also be fit if the distribution is known. Adding more lines improves the fit. The most significant constraints come from using the first line and folding in all lines, including those with critical densities $\lesssim$ the mean of the distribution.}
\label{fig:multiline}
\end{figure*}

So far, we have considered how changes in a single line ratio map to changes in the underlying density distribution. We saw results for different lines in Table \ref{tab:predict} and discussed the qualitative use of multiple lines in Section \ref{sec:ratios}. Modern surveys will often capture many of these transitions \citep[e.g., see][]{BIGIEL16}, so that a pattern of ratios analogous to what we saw in Section \ref{sec:ratios} are available for each source. 

In Figure \ref{fig:multiline}, we explore the use of ensembles of lines to fits the underlying density distribution. We use the model grid from the previous sections, varying $n_0 = 10^2{-}10^4$~cm$^{-3}$, $\sigma =0.6{-}1.2$~dex, and $T_{\rm kin}=15{-}35$~K. Again, we consider pairs of models. A pair of model implies a change in line ratios, dense gas fraction, $n_{\rm med}^{\rm mass}$, $n_0$, and $\sigma$. We draw $1,000$ random pairs of models from the larger model grid. For each drawn model pair, $A$, we imagine that we have observed the line ratio variations implied by this model pair. Then, we search the whole grid for all model pairs that produce the same line ratio variations within a tolerance of $0.05$~dex ($\sim 12$\%). This resembles the typical $\sim 10{-}20\%$ calibration uncertainty for mm-wave observations. These are "good fit" model pairs, that would match the line ratio pattern for A within the uncertainties.

We compare the variation in $f_{\rm dense}$, $n_0$, $\sigma$, and the median $n_{\rm H2}$ for good fit model pairs to those known for $A$. Based on the scatter in these parameters for good-fit model pairs about the true values for $A$, we gauge the rms accuracy with which a given line suite can access the true variations in each quantity.

We repeat this exercise using only the HCN/CO ratio, then using both the HCN/CO and HCO$^+$/CO ratio, then using CO along with all four dense gas tracers, and finally using all lines in our model. For comparison, we measure the scatter across the model grid itself. We show the results for a lognormal distribution with ({\em left}) and without ({\em right}) a tail in Figure \ref{fig:multiline}.

Figure \ref{fig:multiline} illustrates a few key points. First, constraints on the distribution width, $\sigma$, and mean density, $n_0$, are individually weak. The combination of the two produces $n_{\rm med}^{\rm mass}$, and this quantity is better constrained. Fitting using only HCN/CO already improves the accuracy dramatically compared to randomly drawing from the model grid. Adding additional lines further improves the ability to recover the $n_{\rm med}^{\rm mass}$ for lognormal distributions with and without tails. Each additional line helps better constrain this quantity, including folding in the full suite of CO transitions (``All Lines'').

Unsurprisingly, the dense gas fraction is the best-constrained quantity in this exercise. Here, too, additional lines add accuracy, though the gain from additional dense gas tracers is relatively modest, improving the accuracy of the fit by $\sim 0.1$~dex. Again, folding in $^{13}$CO and several low-$J$ CO lines improves our recovery of $f_{\rm dense}$, dropping uncertainties to $\sim 0.1$~dex.

The exact numerical results in Figure \ref{fig:multiline} depend on our adopted model grid, knowing the optical depths of our lines, and other details. But the qualitative conclusions should be robust. The mean density and distribution width are individually less well constrained than $n_{\rm med}^{\rm mass}$, which they combine to set. In the lognormal case, one does some constraint on $\sigma$, but the low dynamic range expected is an issue. The dense gas mass fraction is already constrained well by a single line ratio. This accuracy can be improved by the combination of many lines. Perhaps surprisingly, adding a suite of low-$J$ CO lines can substantially improve the fit.

Note that in practice, the model grid does not represent a suite of equally likely conclusions. Using external knowledge of the temperature or  the distribution width (e.g., from the Mach number gauged via measured velocity dispersions) will improve the ability to infer the full density distribution. More, as discussed above, a key role of multiple dense gas tracers is to provide some robustness to abundance and optical depth variations. That role is not reflected in Figure \ref{fig:multiline}.

\subsection{Dense Gas Conversion Factors}
\label{sec:alpha}

\begin{deluxetable}{lrrr}
\tabletypesize{\scriptsize}
\tablecaption{Dense Gas Conversion Factors \label{tab:alpha}}
\tablewidth{0pt}
\tablehead{
\colhead{Line} & 
\colhead{$\tau \leq 1$} &
\colhead{$1 \leq \tau \leq 3$} &
\colhead{$\tau \geq 3$}\\[2px]
\hline\\[-6px]
\multicolumn{4}{c}{$\alpha_{\rm dense} \pm {\rm scatter}$}\\
\multicolumn{4}{c}{(\acounits)~$\pm$~(dex)}
}
\startdata
HCO$^+$& $1.5 \pm 0.16$~dex & $2.3 \pm 0.23$~dex & $5.3 \pm 0.34$~dex \\
CS & $8.2 \pm 0.11$~dex & $11.4 \pm 0.16$~dex & $23.7 \pm 0.28$~dex \\
HNC & $3.4 \pm 0.08$~dex & $4.8 \pm 0.14$~dex & $9.8 \pm 0.25$~dex \\
HCN & $4.5 \pm 0.06$~dex & $6.3 \pm 0.10$~dex & $12.2 \pm 0.21$~dex
\enddata
\tablecomments{Median and scatter in $\alpha_{\rm dense}$ across the model grid where $n_0 = 10^{2}{-}10^{4}$~cm$^{-3}$, $\sigma = 0.6{-}1.2$~dex, and $T_{\rm kin} = 15{-}35$~K. Here dense gas is defined using a threshold density $n_{\rm thresh} = 10^{4.5}$~cm$^{-3}$. All calculations assume an abundance $X ({\rm mol}) = 10^{-8}$. The quoted scatter does not account for uncertainties in this quantity, which affects the answer linearly. Note that the table quotes linear $\alpha_{\rm dense}$ but logarithmic scatter (in dex).}
\end{deluxetable}

\begin{figure}
\plotone{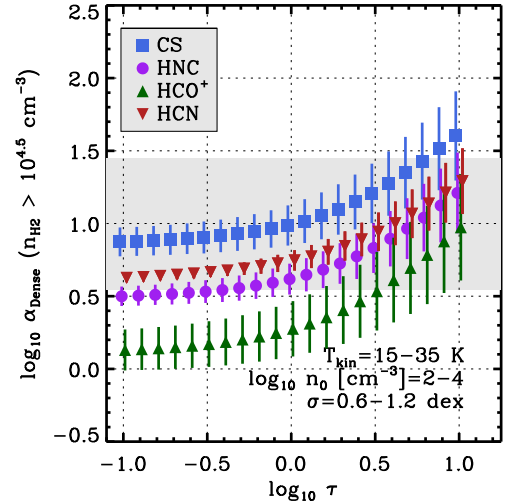}
\caption{Conversion factors relating line emission to dense gas mass for a dense gas threshold of $n_\mathrm{thresh} = 10^{4.5}$~cm$^{-3}$. The gray region shows the range of {\em dense} gas conversion factors for CO, and so represents a control (this much scatter offers little predictive power). The colors show results for HCO$^+$ (purple), HCN (red), HNC (blue), and CS (green), all using our fiducial abundance of $10^{-8}$.}
\label{fig:alpha}
\end{figure}

We have emphasized tracking changes in the gas density via changing line ratios. A more direct approach is to simply translate the flux of HCN, HCO$^+$, or another dense gas tracer into a mass of dense gas. \citet{GAO04B,GAO04} suggest $\alpha_{\rm HCN} \approx 10$~\acounits\ to translate HCN luminosity to the mass of gas at $n_\mathrm{H2} \gtrsim 3 \times 10^4$~cm$^{-3}$. Our calculations yield conversion factors for each line and each grid point. 

How good of an approximation is a fixed $\alpha_{\rm HCN}$ or a fixed $\alpha_{\rm HCO^{+}}$? Considering the same range of densities, distribution widths, and temperatures as above, we calculate the conversion factor relating each dense gas tracer to the dense gas mass for each model. Following the previous section and \citet{GAO04B}, we adopt a threshold of $n_\mathrm{H2} \gtrsim 3 \times 10^4$~cm$^{-3}$ to define dense gas. In this exercise, we consider distributions with and without a power law tail together.

For each line, we consider models with these physical parameters described by a range of optical depths. At each $\tau$, we find the mean and the scatter in $\log_{10} \alpha$, which we plot in Figure \ref{fig:alpha}. As a control, we show the scatter in the dense gas conversion factor for CO~(1-0) as a gray region. We expect (and find) CO on its own to do a poor job of tracing the dense gas. When the scatter in a dense gas conversion factor approaches this value, the utility of that conversion factor is also limited. All conversion factors produced by the model will scale with our adopted abundance, taken as $X ({\rm mol}) = 10^{-8}$ for all lines here. We summarize the results in Table \ref{tab:alpha}.

We calculate $\alpha_{\rm dense,\,HCN}$ between about $3$ and $30$ \acounits . For matched abundances, HCN and HNC yield similar conversion factors. HCO$^+$ produces a lower conversion factor, reflecting its lower effective critical density, and CS~(2-1) has a moderately high conversion factor. In each case, the scatter at fixed optical depth is modest in the optically thin case, but becomes larger when the line becomes optically thick. Most simply, this reflects temperature variations, which will linearly affect $\alpha$ when the line is thick. 

Accounting for the two times higher HCN abundance assumed by \citet{GAO04}, these calculations agree well with their $\alpha_{\rm HCN} \approx 10$~\acounits\ for densities above $\sim 3 \times 10^4/\tau $~cm$^{-3}$. This general agreement belies significant uncertainty. Both the optical depth and the abundance (and thus chemistry) must be known to calculate the conversion factor using this approach. In the optically thick regime, the temperature also plays a large role. To first order, all of these quantities act linearly, so that even a factor of two uncertainty in each (which seems optimistic) implies factor of $\sim 3.5$ uncertainties.

To understand the uncertainty in dense gas conversion factors, it may be helpful to contrast these with the CO-to-H$_2$ conversion factor \citep[see review in][]{BOLATTO13A}. As Figure \ref{fig:beameps} shows, $\beameps$ (and so $\alpha_{\rm CO}$) of CO~(1-0) varies only weakly across a wide range of density distributions, so that $\alpha_{\rm CO}$ does vary strongly as a function of the sub-beam density distribution, all other things held equal. Meanwhile, the assumption of a fixed dynamical state, either marginal boundedness or virialization, has been widely argued and observed for whole clouds \citep[e.g., see][]{BOLATTO13A,HEYER15}. More, the $^{12}$CO lines are usually optically thick, if only because CO represents the dominant carbon reservoir in the well-shielded parts of molecular clouds. In the CO case, the line width of a cloud is set by the cloud dynamical state. The high opacity combines with modest temperature variations to fix the specific intensity in the line at a weakly varying brightness temperature. Then at fixed metallicity, with $\alpha_{\rm CO} \propto \rho^{0.5}~T_{\rm kin}^{-1}$ \citep[e.g.,][and many others]{MALONEY88,NARAYANAN12,BOLATTO13A}. 

Neither high opacity nor a fixed dynamical state can be taken as a given for the gas traced by HCN, HCO$^+$, or similar lines. Unlike CO, these do not represent a main carbon reservoir and their absolute abundances are often quite low. These lines are observed to have some opacity, but not the pervasive optical thickness of $^{12}$CO. Meanwhile, the densities traced by these lines represents a subset of the cloud. For clouds of changing mean density, virial parameter, and with different self-gravitating substructures the dense gas tracers may or may not trace exactly the emission from this tail. Thus, while, \citet{GAO04} argue for a fixed dynamical state for HCN-emitting regions, there does not appear to be a strong {\em a priori} reason to take specifically the HCN-emitting gas to be in a particular, universal dynamical state. 

Instead of assuming a dynamical state for a particular line, the natural place to introduce such a criteria to our model is to couple the column and line width integrated over the power law tail, which could be taken at least to be gravitationally bound. The issue is that this couples the column density of H$_2$ to the line width, again requiring one to adopt an abundance for $N_\mathrm{H2} / \Delta v$ to imply an optical depth in any given line.

Thus, Figure \ref{fig:alpha} and Table \ref{tab:alpha} give approximate conversion factors consistent with previous work. These highlight the systematic variation of $\alpha$ with optical depth. They also give a sense of the relative conversion factors for different lines modulo different abundances. But these calculations remain subject to large systematic uncertainties related to the sub-beam density distribution, opacity, abundance (chemistry), and temperature. Many of these factors can divide out of a relative approach. Our view remains that because $\tau$ is observationally accessible, fixing $\tau$ and comparing line ratios offers a more robust path than invoking an absolute conversion factors.

\subsection{Implications for Isotopologue Studies}
\label{sec:isotopes}

\begin{figure*}
\plottwo{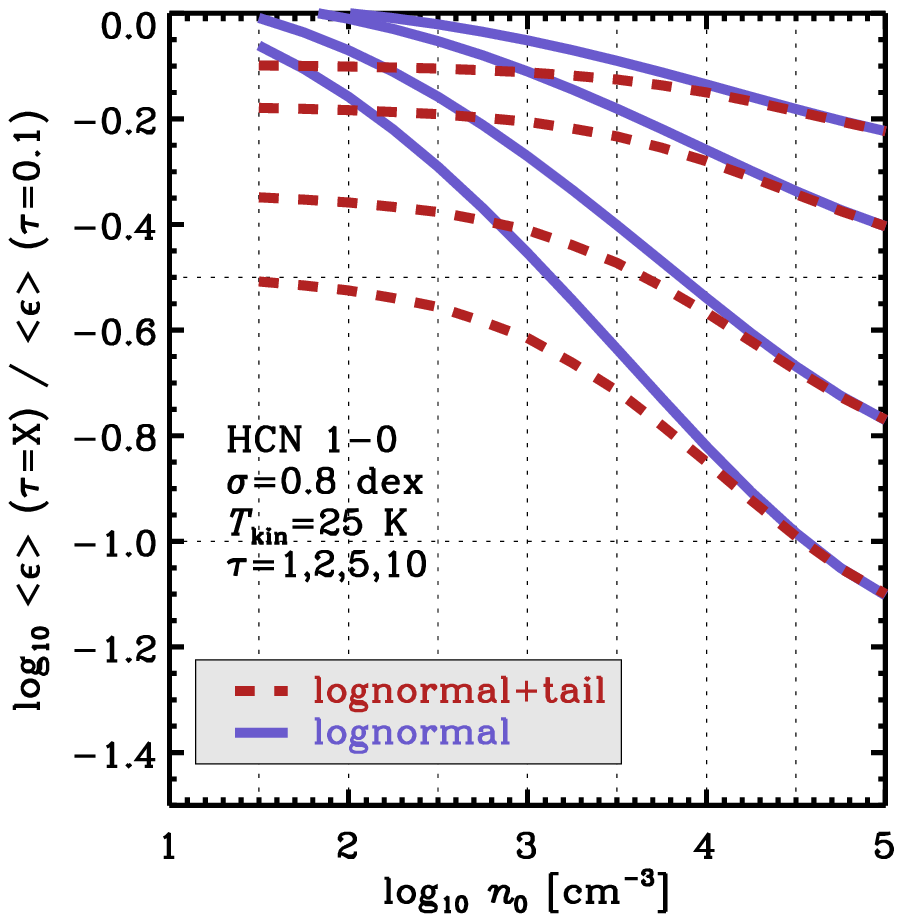}{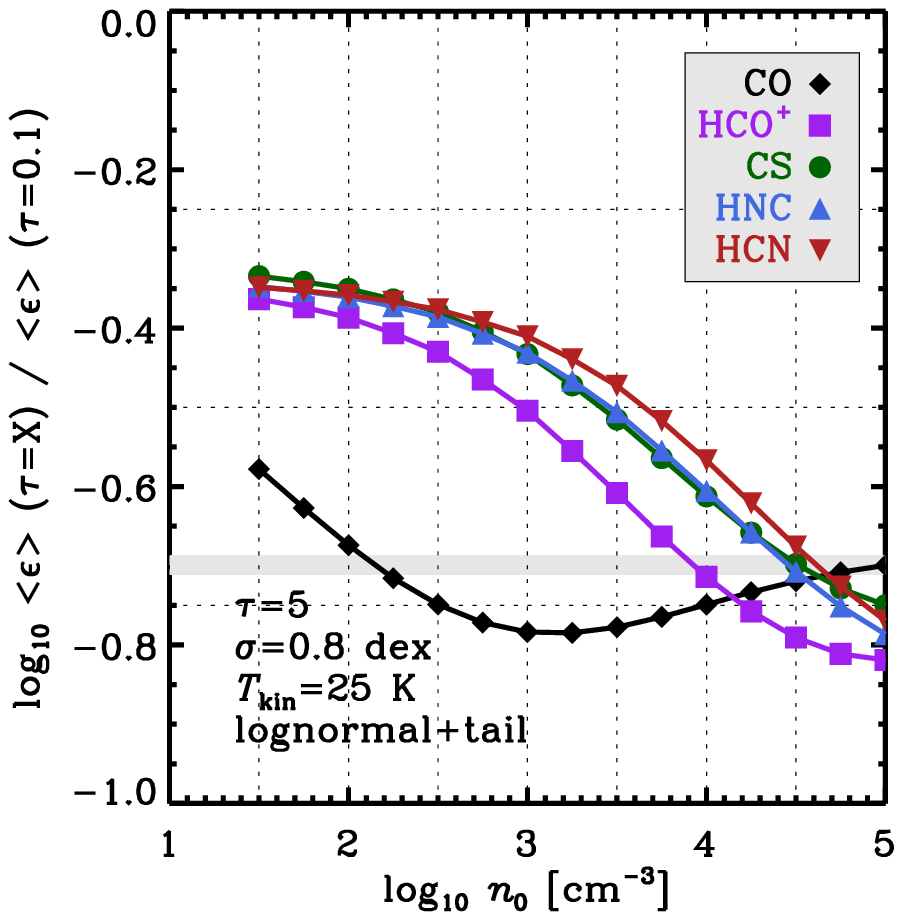}
\caption{Effect of differential excitation on measured isotopologue ratios. Beam averaged emissivity of an optically thick line divided by that of the same line when optically thin. The simple LTE case with $\tau \gg 1$ and LTE predicts a ratio of $\tau^{-1}$. However, taking into account a distribution of densities, emission from the optically thin line is often weaker than one would expect. This results in apparent suppression of the line ratios often used to gauge optical depth, one that must be accounted for to avoid highly biased results in many common cases. An analogous effect should be included to estimate $T_{\rm kin}$ from multi-$J$ observations that integrate over density distributions.}
\label{fig:isotopes}
\end{figure*}

Our model assumes that the optical depth is known, or can be known. The main driver for this assumption is that rare, optically thin isotopologues offer the chance to gauge the optical depth of the main dense gas tracing lines. Alternatively, such observations may constrain the isotopic abundance, which can give clues to recent nucleosynthesis or enrichment patterns as well as cloud chemistry.

The optical depth of a line changes its $\epsilon (n_{\rm H2})$, and thus its sensitivity to gas density. When a distribution of densities exists within the beam, the thin transition will arise from a denser subset of the gas than the optically thick main line. In distributions like those that we consider, this can have the effect of depressing emission from optically thin isotopologues of optically thick dense gas tracers. The effect will be strongest when the mean density of the cloud is lower than the effective critical density of the lines in question, so that density strongly affects emissivity. Thus, we expect this to be a larger effect for HCN or HCO$^+$ than for CO.

Figure \ref{fig:isotopes} shows this effect. The key physics appear in the ratio of the beam averaged emissivity at some optical depth, $\beameps (\tau)$, to the beam averaged emissivity in the optically thin case, $\beameps (\tau \approx 0.1)$. In the limit of $\tau \gg 1$, for local thermodynamic equilibrium and otherwise fixed physical conditions, we expect this ratio to approach $\tau^{-1}$. If the relative abundance of the isotopologues being studied is known, one is optically thin, and one is optically thick, then the abundance ratio is multiplied by this $\tau^{-1}$ to produce the expected line ratio (e.g., CO typically has $\tau \approx 10$ and a $^{12}{\rm C} / ^{13} {\rm C} \approx 50$; therefore typical $^{12}$CO/$^{13}$CO ratios are $\sim 50 \times 10^{-1} \approx 5$).

Figure \ref{fig:isotopes} plots deviations from this simple case due to differential density sensitivities, which also incorporate excitation. In the left panel, we hold all other things fixed, and we compare the emissivity of HCN at $\tau = 1, 2, 5$, and $10$ to HCN at $\tau = 0.1$. For distributions centered at high densities, the expectation of a ratio approaching $\tau^{-1}$ roughly holds. However, as we change the mean density of the distribution, $n_0$, the ratio changes. The sense is that for lower mean densities, the beam-averaged emissivity of the optically thin gas becomes much worse relative to the optically thick gas. This is most extreme for the lognormal case at low mean densities. However, even when a power law tail is present, the effect can be large.

The right panel shows the same plot taking $\tau = 5$ for all four dense gas tracers and $^{12}$CO~(1-0). Although $^{12}$CO shows some differential excitation effects at low densities, the effect here is far more modest. All of the dense gas tracers show significant deviations from the simple expectation for mean densities $n_0 \lesssim 10^3$~cm$^{-3}$, with HCO$^+$ requiring lower densities to show these variations. These differential excitation effects become important when an appreciable amount of gas lies below the critical density of the line in question. This makes isotopologue studies using CO somewhat robust, and HCO$^+$ --- which has an intermediate critical density --- more robust than HCN or HNC, all other things being equal.

Taking the lognormal with a tail, Figure \ref{fig:isotopes} shows that for $\tau \approx 5$, the magnitude of the correction from the simple case is a factor of $\sim 2{-}3$. That is, the ratio of an optically thin tracer like H$^{13}$CN to HCN will be $\sim 2{-}3$ times lower than the expectation for the simple case if $n_0 \lesssim 10^3$~cm$^{-3}$. The difference, due to differential excitation, should not be confused with optical depth or abundance effects.

To apply these results generally, one can take our tabulated calculations and, for matched conditions, compare $\beameps (\tau) / \beameps (\tau \approx 0.1)$ to either the simple expectation ($\sim \tau^{-1}$) or the value found at high densities. This factor represents the part of the line ratio due only to differential excitation and not abundance or optical depth effects.

Optically thin isotopologues remain a crucial way to probe the optical depth of mm-wave lines and isotopic abundance, especially in other galaxies. Indeed, given the importance of optical depth to understand the density-sentivity of the lines that we discuss, more such observations are crucial. However, Figure \ref{fig:isotopes} cautions that sub-resolution density distributions and differential excitation should be borne in mind when interpreting observed ratios. Without taking these into account it would be easy to solve for an optical depth lower than the true value while observing low density gas. Considering isotopologue ratios for HCN, HNC, and HCO$^+$, Jimenez-Donaire et al. (MNRAS submitted) demonstrate the need for such corrections in real measurements. Their synthesis of the literature and new measurements and limits show high HCN/H$^{13}$CN and HCO$^+$/H$^{13}$CO$^+$ ratios in regions where $n_0$ should be $\lesssim 10^3$~cm$^{-3}$. These ratios yield different interpretations in the LTE case and the case described here.

\section{Discussion and Summary}
\label{sec:discuss}

Integrating over a wide range of densities within a single telescope beam is unavoidable when studying molecular gas at extragalactic distances. To date there has been limited effort to interpret spectral line observations in the context of realistic sub-beam distributions, with modeling mostly considering one or two-phase media. To move forward, mm-wave line emission must be modeled taking into account realistic sub-resolution distributions, similar to the treatment of stellar populations in population synthesis or multiple dust populations in IR SED modeling.

In this paper, we consider the interaction of realistic density distributions with the emissivity of common mm-wave transitions. We focus on a suite of transitions in the $85{-}115$~GHz range that are commonly used to trace density. Building on work by \citet{KRUMHOLZ07B}, our model treats emission within a beam as the sum of emission from a collection of one-zone models that share a characteristic optical depth, $\tau$, and temperature, $T_{\rm kin}$, but have a realistic distribution of collider densities, $n_{\rm H2}$.

To implement this, we use the one-zone non-LTE code RADEX \citep{VANDERTAK07} to calculate the emissivity, $\epsilon$, of gas in each line across a range of $n_{\rm H2}$, $T_{\rm kin}$, and $\tau$ giving $\epsilon (n_{\rm H2}, T_{\rm kin}, \tau)$. We calculate emission from a density distribution by combining one-zone models in which the escape probability (via $\tau$) is fixed, while varying the collider density, $n_{\rm H2}$. We implement two density distributions: a pure lognormal and a lognormal distribution that exhibits a power law tail at high densities. Theoretical, numerical, and observational work all commonly invoke this combination of density distributions as a reasonable description of the cold, turbulent gas that produces low-$J$ mm-wave line emission \citep[e.g.,][among many others]{VAZQUEZSEMADENI94,PADOAN02,KRUMHOLZ05,KAINULAINEN09,FEDERRATH13}, though there are some caveats from recent Milky Way work \citep{LOMBARDI15}.

While \citet{KRUMHOLZ07B} and the closely-related paper by \citet{NARAYANAN08} mainly focused on star formation scaling relations, we are interested in the use of mm-wave line emission to trace gas density within a telescope beam. Because of the small physical scales involved, spectroscopy offers almost the only way to access the small-scale density distribution across a wide range of extragalactic environments. Despite the faintness of the high effective density lines like HCN \mbox{(1-0)}, HCO$^+$ \mbox{(1-0)}, and CS \mbox{(2-1)}, the Atacama Large Millimeter/submillimeter Array, the Green Bank Telescope, and the IRAM telescopes are now able to regularly observe these transition across the disks of normal galaxies \citep[e.g.,][Gallagher et al. (in prep.), among others]{KEPLEY14,USERO15,BIGIEL16}.

The paper includes a large set of tabulated results available as online-only material (Tables \ref{tab:neff} and \ref{tab:dist_tab}). These report:

\begin{enumerate}
\item The minimum collider density, $n_{\rm H2}$, at which each line achieves $95\%$ of its maximum emissivity for a given $T_{\rm kin}$ and $\tau$ in a one-zone model (Table \ref{tab:neff}). This is closely related to the effective critical density \citep[e.g.,][and references therein]{SHIRLEY15}. Having such values tabulated for the non-LTE case, cast directly in terms of emissivity and calculated across the transition from optically thin to optically thick lines, may be useful to readers.

\item The beam-averaged emissivity, \beameps , and median density for emission in each studied transition for a range of $T_{\rm kin}$, $\tau$, and density distributions comparable to those found across the local galaxy population (Table \ref{tab:dist_tab}). From these, line ratio patterns can be constructed for any adopted set of optical depth and temperature.
\end{enumerate}

\subsection{Using Line Ratios to Trace Density}
\label{sec:disc_rats}

We examine the ability of mm-wave line ratios to trace the sub-beam density distribution. For these purposes, the interaction between the emissivity, $\epsilon (n_{\rm H2}, T_{\rm kin}, \tau)$, and the density distribution, $P(n_{\rm H2})$, is crucial. From a mixture of one zone models and combinations that model a density distribution, our conclusions include:

\begin{enumerate}

\item Gas can still emit effectively at densities well below its effective critical density. That is, the emissivity, $\epsilon (n_{\rm H2})$, of a mm-wave transition still has high values for at least a decade in $n_{\rm H2}$ below the density of peak emissivity (see Figure \ref{fig:emis_vs_dens}).

\item The flux that emerges from each density in the cloud is the product of the mass at that density and the emissivity at that density. For realistic distributions, $P(n_{\rm H2})$ drops with increasing $n_{\rm H2}$ near the critical density of the dense gas tracers. This often leads to the case where the density distribution and emissivity exert competing effects: there is more mass at lower densities but gas emits better at high densities (see Figure \ref{fig:pdf_illus}).

\item For a pure lognormal distribution, the interaction between $\epsilon (n_{\rm H2})$ and $P(n_{\rm H2})$ leads to large variations in the median density producing emission from high density tracers. We illustrate this for HCN \mbox{(1-0)}, and a similar case holds for other lines. In cases of low mean density, $n_0$, and narrow distribution width, $\sigma$ (expected for low Mach number), the median density of gas producing HCN emission can fall well below the critical density of HCN. When this happens the overall emissivity of the gas in HCN also drops. Thus we for regions in which modest density gas produces most HCN emission, HCN/CO and similar line ratios will also be low (see Figure \ref{fig:neff_dist}).

\item The presence of a power law tail in the density distribution tends to suppress variations in the median density for emission. Such tails are expected for self-gravitating gas and lack a preferred scale. As a result, they produce much weaker variations in the median density for emission. For example, HCN emission from a distribution with our fiducial power law tail almost always arises from $\sim 10^4{-}10^5$~cm$^{-3}$ gas (see Figure \ref{fig:neff_dist}).

\end{enumerate}

The sub-beam density distributions also affects the beam-averaged emissivity, \beameps , so that some lines emit better for certain density distributions.

\begin{enumerate}
\setcounter{enumi}{4}

\item Variations in \beameps\ are strongest for the dense gas tracers, which have high \beameps\ when more dense gas is present. \beameps\ remains more nearly constant for the low-$J$ CO lines, reflecting their common use as total gas tracers. As a result of these differential variations in \beameps , line ratios have the power to capture changes in the density distribution within the beam (see Figures \ref{fig:beameps} and \ref{fig:linerat}).

\end{enumerate}

The sense of these variations is that lines with effective critical densities low compared to the mean density, like the low-$J$ CO transitions, vary little in their emissivity. Transitions with effective critical densities high compared to the mean density see their emissivity vary strongly as the density distribution changes. Thus, ratios like HCN-to-CO have the power to probe the fraction of dense gas or the mean gas density within a beam.

The exact value of \beameps , or any individual line ratio, depends on our adopted abundances. These are often poorly known. To help circumvent this concern, we highlight the power of line ratio variations. 

\begin{enumerate}
\setcounter{enumi}{5}

\item In the case that two regions have different density distributions and the abundances remain fixed (but also unknown), the measured variation in line ratios has the power to trace the change in density distribution within the beam, independent of the absolute adopted abundance (see Figure \ref{fig:linerat}).

\item These line ratio variations will be strongest for lines with the highest effective critical densities. Thus, the signature of a changing density distribution is a ``flaring'' pattern of line ratios when the lines are sorted as a function of increasing critical density (see Figure \ref{fig:linerat}). 

\item The shape of this flare reflects the high end form of the density distribution. For a self-similar distribution like a power law, the ratios of all lines on the power law tail relative to CO will vary in lock-step. For a downward curving, steep distribution like a lognormal, line ratio variations will increase in strength with increasing critical density (see Figure \ref{fig:linerat}).

\end{enumerate}

Beyond accessing the shape of the high density tail, line ratio patterns allow some prospect to control for abundance variations. If several high density tracers are observed, then consistency (or lack thereof) among their variations can help distinguish changing abundances from changing density. 

\begin{enumerate}

\setcounter{enumi}{8}
\item We also examine the impact of temperature variations on changing line ratio patterns. In general, $T_{\rm kin}$ variations exerts a weaker influence on line ratios than changing density because they affects all lines, including those with with low critical densities. However, $T_{\rm kin}$ does still have a some impact on the line ratios because it lowers the effective critical density of the dense gas tracers (see Figure \ref{fig:linerat_temp}).

\end{enumerate}

We quantify how changes in the line ratio map to variations in the dense gas mass fractions, $f_{\rm dense}$, and $n_{\rm med}^{\rm mass}$. Here, the shape of the high end of the density distribution plays a large role. 

\begin{enumerate} 
\setcounter{enumi}{9}
\item When a power law tail describes the distribution at high densities, HCN/CO, HCO$^+$/CO, and similar ratios trace variations in $f_{\rm dense}$ with approximately linear slope. This remains true for most reasonable definitions of ``dense'' gas. That is, if a power law tail is present and approximately universal, then the interpretation of dense gas tracers may be relatively straightforward (see Figure \ref{fig:lrat_fdense}).

\item On the other hand, if the density obeys a steep, curving distribution like a lognormal, the interpretation of line ratio variations becomes more complex. We calculate scaling relations relating variations in HCN/CO and other line ratios to variations in $f_{\rm dense}$, but caution that these are nonlinear and depend on the adopted threshold density. Still within a factor of $\sim 2{-}3$ scatter, these ratios can capture variations in the dense gas mass fraction and $n_{\rm med}^{\rm mass}$ (see Figures \ref{fig:lrat_fdense} and \ref{fig:lrat_nmed}). 
\end{enumerate}

Given strong priors on the Mach number or mean density, the grid of models that we consider could be narrowed considerably. In this case, the inference of changes in $f_{\rm dense}$ and $n_{\rm med}^{\rm mass}$ from line ratio variations can be even more precise. 

Current surveys often capture a suite of line ratios. We explore the use of an ensemble of ratios to estimate changes in the sub-resolution density distribution. We do this by directly fitting line ratio variations to our grid of models. This approach can be implemented generally using the online tables included with the paper. 

\begin{enumerate} 
\setcounter{enumi}{11}
\item We show that including multiple lines improves the accuracy with which variations in the sub beam density distribution are recovered. Variations in the mean density, $n_0$, and distribution width, $\sigma$, are, in general, less well recovered than their combination, $n_{\rm med}^{\rm mass}$. Changes in the dense gas fraction, $f_{\rm dense}$ tend to be better recovered than $n_{\rm med}^{\rm mass}$ (see Figure \ref{fig:multiline}).

\item Including more lines in the fit steadily improves the accuracy with which the model variations are recovered. The best fit comes from including all dense gas tracers and multiple tracers of lower density gas, here meaning the low-$J$ and optically thin CO lines (see Figure \ref{fig:multiline}).
\end{enumerate}

\subsection{Differential Excitation and Isotopologue Studies}
\label{sec:disc_iso}

Our model specifies the optical depth rather than the abundance. One motivation for this is that the optical depth of mm-wave transition can be constrained by observations. The most direct way to do this is to pair observations of the main tracers with optically thin isotopologues.

\begin{enumerate} 
\setcounter{enumi}{13}
\item We show that in the presence of the sub-beam density distributions considered by our analysis, differential excitation between the main line and the optically thin isotopologue complicates this analysis. Even treating the two molecules as identical, the optically thin line has different $\epsilon (n_{\rm H2})$. This, convolved with the sub-beam density distribution, can affect the line ratio. If one interprets observations of the optically thin tracer without taking these effects into account, the optical depth inferred can be biased by a factor of $\gtrsim 2{-}3$ (see Figure \ref{fig:isotopes}).

\item The key physics behind this effect are captured in the ratio $\beameps (\tau) / \beameps (\tau = 0.1)$ for fixed $T_{\rm kin}$ and density distribution. This is the ratio of beam averaged emissivity of a molecule at some optical depth to the beam averaged emissivity for that molecule when optically thin. We show this factor for some representative cases, and it can be calculated from our tabulated model grid. When a power law tail is present and the mean density is low compared to the critical density of the thin line, we arrive at the factor of $2{-}3$ connection mentioned above. For a pure lognormal centered at low mean density, the effects can be even more extreme (see Figure \ref{fig:isotopes}).

\item Although we do not focus on temperature in this paper, a related approach to differential excitation should be used to infer temperatures from line ratios.  $\epsilon (n_{\rm H2})$ varies with $T_{\rm kin}$ and with transition; e.g., see the case for CO in Figure \ref{fig:emis_vs_dens}. When using multi-$J$ HCN or HCO$^+$ observations to constrain $T_{\rm kin}$, it will be important into account the convolution of $\epsilon (n_{\rm H2}, T_{\rm kin})$ and $P (n_{\rm H2})$ in order to constrain $T_{\rm kin}$.
\end{enumerate}

\subsection{Dense Gas Conversion Factors}
\label{sec:disc_conv}

We emphasize differential measurements, because these may help control for an unknown absolute abundance scale. However, it will still often be of interest to discuss absolute masses or mass fractions of dense gas. To this end, we examine the dense gas ``conversion factors'' implied by our models (see Figure \ref{fig:alpha}). These are formally the ratio \beameps\ to $f_{\rm dense}$, and give the factor by which line luminosity should be multiplied to calculate the dense gas mass.

Our derived conversion factors depend on the optical depth, abundance, $T_{\rm kin}$, and the density distribution. This renders them substantially uncertain. The overall normalization that we find agrees with previous work for matched assumptions. But we emphasize that arguments based on a fixed dynamical state at a particular density inside a cloud remain aggressive. Our calculations assume a fixed abundance for all four dense gas tracers. In this case at fixed $\tau$, $\alpha_{\rm HCO^+} < \alpha_{\rm HNC} \approx \alpha_{\rm HCN} < \alpha_{\rm CS}$.

\subsection{Next Steps}
\label{sec:nextsteps}

We have presented a framework for interpreting mm-wave line ratio observations. A number of natural next directions recommend themselves.

{\em Observations:} First and foremost, we need a large set of observations targeting individual regions in other galaxies. These should include transitions that sample a range of densities, similar to the line suite considered in this paper. With such a large database, it will be possible to place empirical constraints on the shape of the high density PDF as well as identify which tracers show coherent variations and which show signatures of strong abundance variations.

Ideally, these observations can be paired with high resolution imaging of ISM structure. Such observations have the prospect to constrain the mean density and Mach number at the scale of whole clouds \citep[e.g.,][]{LEROY16}. Such external constraints on $n_0$ and $\sigma$ offer the prospect to validate the models proposed here. Then, if the models are validated, they offer the prospect to place priors on the model grid when it is applied to individual regions.

We have taken the optical depth to be known, or at least observable. However, the isotopologues of the dense gas tracers discussed here are very faint. Their emission is further suppressed by the differential excitation effects discussed here. As a result constraints on the optical depth of the dense gas tracers are actually quite weak to date. Systematic observation of H$^{13}$CN, H$^{13}$CO$^+$, $^{13}$CS, etc. in the disks of normal galaxies will be essential to break a key degeneracy in the model.

Fortunately, current mm-wave facilities can make many of these key observations. ALMA, the Green Bank Telescope (GBT), the IRAM telescopes, and other modern facilities have the ability to survey dense gas tracers and cloud properties across nearby galaxies. Several recent, ongoing, and planned surveys promise to expand our knowledge of dense gas tracers across nearby galaxies \citep[e.g.,][Gallagher et al., in prep.; and upcoming efforts using ARGUS on the GBT]{USERO15,BIGIEL16}. Meanwhile, ALMA and NOEMA are rapidly expanding our knowledge of cloud-scale ISM structure, often in overlapping targets \citep[e.g.,][]{HUGHES13A,LEROY16}.

Despite good prospects for the next few years, many key lines remain very faint. Systematic surveys of optically thin isotopologues (e.g., Jimenez Donaire et al., MNRAS submitted) or imaging of dense gas tracers in normal galaxies at high physical resolution \citep[e.g.,][]{ROSOLOWSKY11} will remain very challenging in normal star-forming galaxies beyond the Local Group. Similarly, sensitivity considerations limit the suite of available lines for extragalactic studies compared to those used in the Galaxy. The long term prospects for such studies may require an increase even beyond the capabilities of ALMA. One natural facility to enable such studies would be a Next Generation Very Large Array optimized for mm-wave studies. Large receiver arrays on large single dish telescopes offer another promising direction.

{\em Power Law Tails:} A recurring result of our analysis is that the shape of the density distribution at high $n_{\rm H2}$ exerts a large influence. The results for a power law tail and a pure lognormal distribution differ substantially. A main way to improve the interpretation of high density tracers is to resolve the question of how common a power law tail is, whether it has a universal slope, and the threshold density for its onset.

Most current extragalactic surveys target large areas of star-forming galaxies, often selected for local or global activity. We expect such locales to harbor self-gravitating gas that serves as the immediate fuel for star formation. In that sense, we do have a basic expectation that a power law tail should be present at some level in most star-forming regions. We do not, however, have a strong expectation for the universality of the slope of this tail or the density at which it becomes the dominant contributor to the PDF. Our power law models show an optimistic case, in which these factors do not vary much from beam to beam.

Galactic studies have made large progress identifying power law tails in column density distributions and linking these to environment in the cloud and Galaxy \citep[e.g.,][]{ABREUVICENTE15,STUTZ15}. From some combination of these studies, spectral surveys and highly resolved observations of the nearest galaxies, we can hope to understand a characteristic slope, typical range of variation, and condition for onset of a power law tail. For modeling distributions in other galaxies, an approach like that of \citet{DRAINE07B} may be most appropriate, in which the strength of a power law tail relative to the lognormal acts as a free parameter in the model.

\acknowledgments We thank the anonymous referee for a constructive report that improved this paper. This work was carried out as part of the ``Star Formation and Feedback in Nearby Galaxies'' (SFNG) collaboration. The work of AKL is partially supported by the National Science Foundation under Grants No. 1615105 and 1615109. AU acknowledges support from Spanish MINECO grants AYA2012-32295 and FIS2012-32096. JMDK gratefully acknowledges financial support in the form of an Emmy Noether Research Group from the Deutsche Forschungsgemeinschaft (DFG), grant number KR4801/1-1. AH acknowledges support from the Centre National d'Etudes Spatiales (CNES). The National Radio Astronomy Observatory is a facility of the National Science Foundation operated under cooperative agreement by Associated Universities, Inc. ER is supported by a Discovery Grant from NSERC of Canada. G.B. is supported by CONICYT/FONDECYT, Programa de Iniciacion, Folio 11150220. ES acknowledge financial support to the DAGAL network from the People Programme (Marie Curie Actions) of the European Union's Seventh Framework Programme FP7/2007- 2013/ under REA grant agreement number PITN-GA-2011-289313.

\bibliography{akl}

\end{document}